\documentclass[twocolumn]{aastex63}

\usepackage{bm}
\shorttitle{Energy dependent radiation properties in pulsar dissipative magnetospheres}
\shortauthors{Yang \& Cao}
\begin{document}

\title{Exploring the energy-dependent radiation properties in dissipative magnetospheres with Fermi pulsars}

\author{Xiongbang Yang}
\affiliation{Department of Astronomy, Yunnan University, Key Laboratory of Astroparticle Physics of Yunnan Province, Kunming 650091, Yunnan, P. R. China; xbyang@mail.ynu.edu.cn;xiongbangyang@hotmail.com}
\author{Gang Cao}
\affiliation{Department of  Mathematics, Yunnan University of Finance and Economics, Kunming 650221, Yunnan, P. R. China;}

\begin{abstract}
    The equatorial current sheets outside the light cylinder(LC) are  thought as the promising site of the high energy emission based on the results of the recent numerical simulations. We explore the pulsar light curves and energy spectra by computing the curvature radiation based on the FIDO magnetospheres. The FIDO magnetospheres with a near force-free regime inside the LC and a finite but high conductivity outside the LC are constructed by a spectral algorithm. The pulsar high energy emission properties are explored by integrating the trajectories of the test particles under the influence of both the accelerating electric field and the curvature radiation losses. As an application, we compare the predicted energy-dependent light curves and energy spectra with those of the Crab and Vela pulsars published in Fermi 2PC catalog. We find that the observed characteristics of the light curves and energy spectra from Crab and Vela pulsars can be well reproduced by the FIDO model.
\end{abstract}


\section{Introduction}
Pulsars are rapidly rotating and highly magnetized neutron stars, which were discovered for almost half a century \citep{hew68}. More than 230 $\gamma$-ray pulsars have been detected by Fermi-LAT, in which 117 pulsars are listed in the Second Fermi Pulsar Catalog (2PC) \citep{abd13}. The observed typical features of the gamma-ray pulsars are the double-peaked light curves with significant bridge emission and off-peak emission, and the first peak are commonly lagged the radio pulse by several rotation periods. The pulsar $\gamma$-ray light curves can be used to constrain the location of particle acceleration and radiation mechanisms in the magnetosphere. The high-quality phase-averaged, and phase-resolved spectra provide us the valuable information to study the  pulsar physical mechanisms in the magnetosphere. However, the precise locations where the particles are accelerated and the mechanisms how they are accelerated are still unclear.

In the early stage of the pulsar study, the pulsar magnetosphere was usually treated as an vacuum retarded dipole (VRD) \citep{deu55}. Based on the VRD field geometry, the polar cap (PC) \citep[e.g.,][]{rud75,har78,dau82,dau96}, slot gap (SG) \citep[e.g.,][]{as79,ar83,mus03,mus04,dyk03,dyk04,bai10a}, outer gap (OG) \citep[e.g.,][]{che86a,che86b,zc97,ya97,che00,zc00,zc01,bai10a} models were proposed to model the pulsar light curves and energy spectra. The VRD field can  produce an accelerating electric field on the the surface of the star. Such accelerating electric field  pulls the particles from the neutron star surface to fill the magnetosphere by the pair cascades,  these particles short out the electric field to form the Force-Free (FF) magnetosphere \citep{gol69}.

Following the pictures of \citet{gol69}, \citet{sc73} deduced the  well-known pulsar equation expressed by the poloidal magnetic flux for an aligned rotator. Many attempts were made to solve the pulsar equation by several groups \citep[e.g.,][]{mi73a,mi73b,en74}. Until the year of 1999, \citet{con99} firstly solved the pulsar equation by a iterative algorithm and obtained the CKF solution.
The CKF solution  was further explored by many groups \citep[e.g.,][]{gru05,con05,ko06,mc06,tim06,yu11,par12,cao16a}. The CKF solution  consists of  the closed region extending to the LC, the open region with asymptotically monopolar magnetic field lines, and the equatorial current sheet outside the LC.

The time-dependent simulation of the force-free pulsar magnetospheres for the oblique rotator was firstly performed based on the the finite-difference time-domain (FDTD) approach by \citet{spi06}. And then the algorithm of \citet{spi06} was improved by \citet{kal09} by implementing the non-reflecting absorbing boundaries so that a final steady-state  solution can be reached when evolving many stellar periods. The pseudo-spectral method was also developed to simulate the 3D FF magnetosphere \citep{pet12,par12,cao16a}. All these simulations converge to a similar  CKF magnetosphere with an equatorial current sheet outside the LC. The FF approximate is also extended to the full magnetohydrodynamic regime that the plasma inertia and pressure are take into account \citep{tec13}  and to the general-relativistic regime that taken into account the time-space curvature and frame-dragging effects \citep{pet16,ca18}.
The force-free solution provide different field structure compared to the vacuum one. The  pulsar light curves are also explored by assuming the location of the accelerating zone based the force-free field \citep[e.g.,][]{con10,bai10b,har15,chen20}.

The force-free solution is non-dissipative and thus preclude the production of the pulsed emission in the magnetosphere. More realistic magnetosphere should allow the local dissipative to accommodate the production of radiation in some regions. Therefore, a resistive magnetosphere was proposed to model the pulsar magnetosphere by involving a macroscopic conductivity parameter \citet{kal12b,li12,cao16b}. The resistive solution ranges from the VRD to FF magnetospheres with increasing conductivity and the dissipative region appears in the equatorial current sheet outside the LC for the high conductivity \citep[e.g.,][]{cao16b}. The resistive magnetosphere were also used to study the high-energy phenomena of pulsar \citep{kal12b,kal14,bra15,cao19}. Recently, the particle-in-cell (PIC) method was used to model the pulsar magnetosphere by self-consistently treating the  particle motions and the electromagnetic fields \citep[e.g.,][]{chen14,ph14,be15,ce15,kal18,bra18}. The pulsar light curves are explored by including the radiation reaction in the PIC code \citep[e.g.,][]{ce16,kal18,ph18}.

In the previous works, the $\gamma$-ray light curve in the dissipative magnetosphere-the FIDO magnetosphere with near FF regime inside the LC and finite conductivity outside the LC is produced by collecting the bolometric luminosity from all emitting particles \citep{kal14,cao19}. However, these studies did not compute the curvature spectrum from the individual particles and the light curves are only produced by collecting the bolometric luminosity from all the emitting particles. \citet{bra15} explored the impact of the $\sigma$ parameter on the light curves and spectra in the FIDO magnetospheres by using an approximate expression for the accelerating electric field. Recently, \citet{kal17} further refined the FIDO models and calculated the curvature radiation spectra using the realistic accelerating fields given by the models.

In this paper, we further explore the $\gamma$-ray energy-dependent radiation patterns through extending the study of \citet{cao19} by computing the curvature spectra from the emitting particles. In section \ref{sect-DM}, we describe the dissipative magnetosphere model. In section \ref{sect-PIT}, how the test particles are injected and tracked. In section \ref{sect-MCR} and \ref{sect-SM}, we elaborate the modeling of the curvature spectra, and how the sky maps and light curves are produced. In section \ref{sect-app}, the FIDO magnetospheres are applied to the Crab and Vela pulsars, their sky maps, light curves, and spectra are produced. Finally, the conclusions and discussions are listed in section \ref{sect-CD}.

\section{The Dissipative Magnetosphere}\label{sect-DM}

A dissipative magnetosphere can be obtained by solving the time-dependent Maxwell equations
\begin{eqnarray}
\frac{1}{c}{\partial {\bf B}\over \partial t}&=&-{\bf \nabla} \times {\bf E}\;,\\
\frac{1}{c}{\partial  {\bf E}\over \partial t}&=&{\bf \nabla} \times {\bf B}-\frac{4\pi}{c}{\bf J}\;,\\
\nabla\cdot{\bf B}&=&0\;,\\
\nabla\cdot{\bf E}&=&4\pi \rho\;,
\end{eqnarray}
where $\bf B$ is the magnetic field, $\bf E$ is the electric field, $\rho$ is the charge density, and $\bf J$ is the current density. In the dissipative magnetosphere, the current density {\bf J} ensuring the closure of the system is defined as a form of the Ohm's law by the local electromagnetic fields with a conductivity parameter \citep{kal14,cao16b}
\begin{eqnarray}\label{eq-current}
{\bf J}= c\rho {{\bf E}\times{\bf B}\over B^2+E^2_{\rm 0}}+\sigma {\bf {E}}_{\|}\;,
\end{eqnarray}
where the first term in equation (\ref{eq-current}) is the drift velocity perpendicular to the magnetic fields, the term $E_{0}$ in the denominator ensures drift velocity to be subluminal and satisfies following conditions: $B^2_{0}-E^2_{0}={\bf B}^2-{\bf E}^2$, $E_{0}B_{0}={\bf E}\cdot {\bf B}, \quad E_{0}\geq0$. The second term, ${\bf {E}}_{\|}={\bf E}\cdot{\bf B}/B$  is the accelerating electric field parallel to the magnetic field, which is consistently controlled by the conductivity parameter $\sigma$.

In this paper, the FIDO magnetospheres with near FF regime (the conductivity fixed as $60\,\Omega$) inside the LC and a dissipative regime of a finite conductivity $\sigma$ outside the LC are constructed by the spectral method, where any $E_{||}$ component within the LC are discarded \citep[see also][]{kal14,cao19}.
The neutron star is treated as a perfect conductor with a magnetic moment in the center of the star. The boundary condition on the star surface is enforced by a co-rotating electric field ${\bf E}= - ({\bf \Omega} \times {\bf r})\times{\bf B}/c$. We evolve the Maxwell equations for a series of magnetic inclination angles $\alpha$ ranging from $15^\circ$ to $90^\circ$ with a interval of $5^\circ$. The conductivities $\sigma$ outside the LC  are set to 0.3\,$\Omega$, 3\,$\Omega$, 10\,$\Omega$, and 30\,$\Omega$. The computational domain extends from the star surface $r_{\star}=0.2 \, {R_{\rm LC}}$ up to $r_{\rm max}=3 \, {R_{\rm LC}}$. A better accuracy can be obtained with the resolution of $N_r\times N_\theta \times N_\phi = 128\times64\times128$. Moreover, the magnetospheres are obtained after several pulsar spin periods to reach the final stable state. The ${\bf E}_{||}$ values outside the LC provided by the FIDO solutions are used to compute the  curvature radiation. This is very different from the ones used in \citet{kal14} and \citet{bra15}, in which they used an approximate expression to produce the accelerating electric field based on the corresponding force-free solutions.

\section{Particles injection and acceleration}\label{sect-PIT}
The realistic pulsar magnetospheres are filled with abundant electron/positron pairs, which are accelerated by the parallel electric fields to relativistic velocity and radiate the $\gamma$-ray photons. In order to imitate the behavior of the electron/positron in the dissipative magnetospheres, a set of $\sim 1.5 \times 10^6$ electron/positron pairs with small initial Lorentz factors (${\gamma_{\rm ini}} \lesssim 100$) are randomly ejected from the PCs. We track the test particles from the neutron surfaces to $2.5 R_{\rm LC}$ to produce the pulsar light curves and spectrum by including both the accelerating electric field and curvature radiation losses. The trajectory of particles \citep{kal14,cao19} in the inertial observe frame (IOF) is given by the local electromagnetic fields
\begin{eqnarray} \label{eq-velocity}
{\bf v}\equiv \frac{d{\bf x}}{dt} = \left( \frac{ {\bf E} \times {\bf B} }{B^2+E^2_{0}} + f \frac{\bf{B}}{B} \right) c \;,
\end{eqnarray}
where the first term in equation (\ref{eq-velocity}) is the drift velocity, while the second term is the velocity component parallel to the magnetic fields. The sign and the value of the scalar factor $f$ is determined by setting $v \simeq c$ and ensure that the particle motions are always outward.

\section{The modeling of the curvature radiation}\label{sect-MCR}

Once the particle trajectories are determined, the Lorentz factors $\gamma$ of the radiating particle along each trajectory can be calculated by integrating the following expression
\begin{eqnarray}\label{eq-gamma}
\frac{d\gamma}{dt}=f\frac{q_{\rm e}c E_{\|}}{m_{\rm e}c^2}- \frac{2q^2_{\rm e} \gamma^4}{3R^2_{\rm CR}m_{\rm e}c},
\end{eqnarray}
where the first term in equation (\ref{eq-gamma}) is the energy gain rates of the particles due to the accelerating electric field, and the second term is the energy loss rates due to CR reaction. $q_{\rm e}$ and $m_{\rm e}$ are the electron charge and rest mass, and $E_{||}$ are the component of accelerating electric fields provided by the solutions themselves. $R_{\rm CR}$ is the local curvature radius at each point of the trajectory, which is calculated in the IOF by
\begin{eqnarray}\label{eq-Rcr}
R_{\rm CR}=\frac {dl}{d\theta}\;,
\end{eqnarray}
$dl$ is the segment length along the particle trajectory, while $d\theta$ is the angle between two adjacent velocities. The expression of the curvature radius used in the paper is essentially the same as the one given by \citet{har15}, but different forms. Recently, similar approach to determine the curvature radius is also performed by introducing the differential geometry Frenet-Serret equations to express the trajectories of the particles by \citet{vig20} .
The equilibrium $\gamma_{\rm L}$ values balanced by the acceleration of $E_{||}$ and the CR loss can be obtained in the
equilibrium of radiative reaction region ($d\gamma/dt=0$)by
\begin{eqnarray}
\gamma^4_{\rm L}=\frac{3f E_{||} R^2_{\rm CR}}{2q_{e}},
\end{eqnarray}
we also note that the $\gamma_{\rm L}$ values are weakly affected by $E_{||}$ and $R_{\rm CR}$ when the equilibrium states are reached.

The energy spectrum of curvature radiation from a single particle at each radiating location $r$ with the Lorentz factor $\gamma$ is calculated by integrating the expression \citep{tang08},
\begin{eqnarray}
F(E_{\gamma},r)=\frac{\sqrt{3} e^2 \gamma}{2 \pi \hbar R_{CR} E_{\gamma}}F(x)\;,
\end{eqnarray}
where  $x=E_{\gamma}/E_{\rm cur}$, $E_{\gamma}$ is the energy of emitting photon, $E_{\rm cur}=\frac{3}{2}c\hbar \frac{\gamma^3}{R_{\rm CR}}$ is the characteristic energy of
the curvature radiation photon, and the function $F(x)$ is defined as
\begin{equation}
F(x)=x\int_{x}^{\infty}{K_{\rm 5/3}}(\xi)\;d\xi,
\end{equation}
where $K_{5/3}$ is the modified Bessel function of order $5/3$. The function $F(x)$ is calculated by using the approximate expression given by \citet{ah10}.
In fact, we only sample $\sim 1.5 \times 10^6$ particles from the stellar surface, which can not reflect real particle numbers in the pulsar magnetospheres. Therefore, we weight the individual flux of curvature radiation by the surface charge density $\rho_{\rm s}$.
\begin{figure}
\begin{tabular}{c}
\hspace*{-1cm}
\includegraphics[width=8.5 cm,height=7 cm]{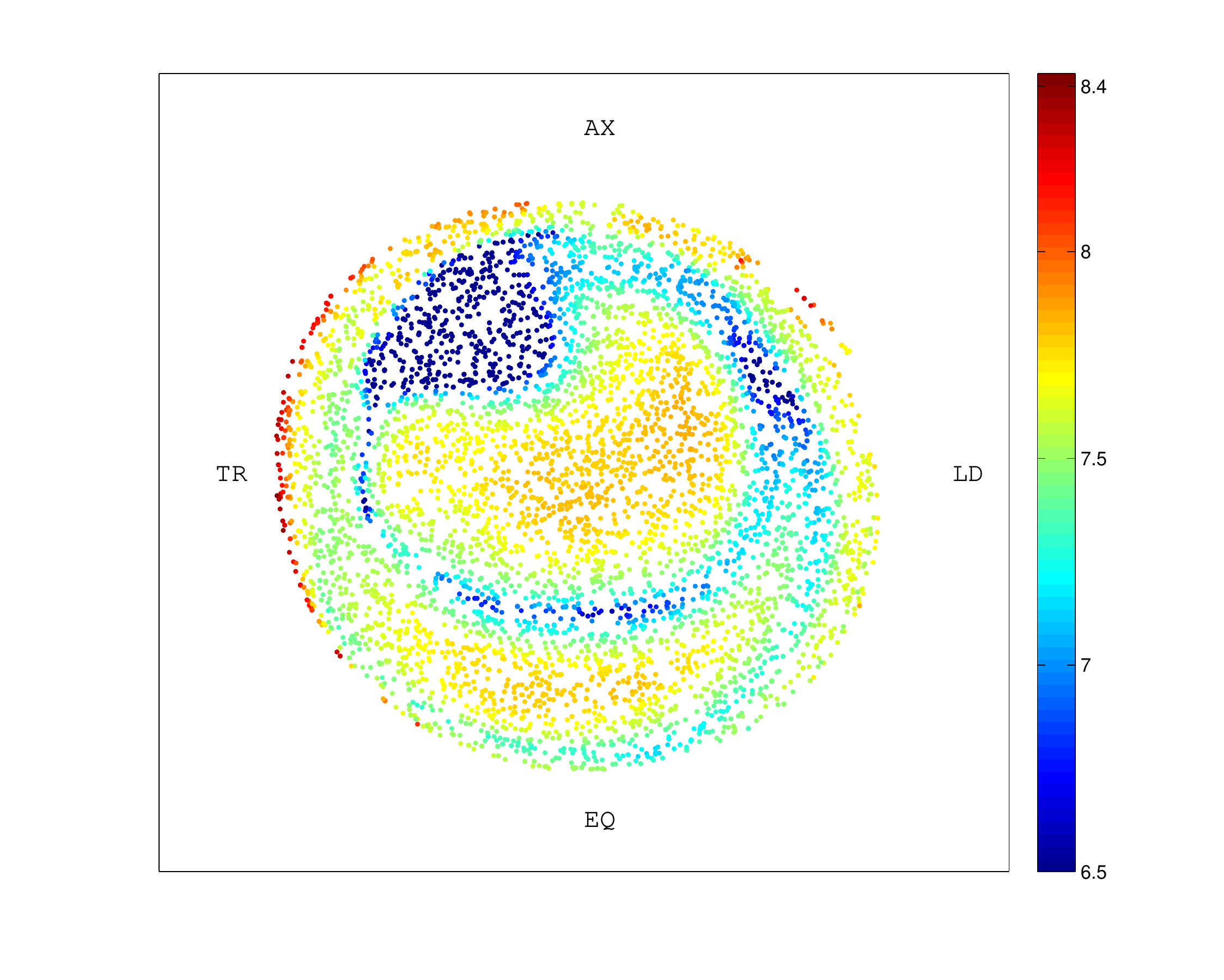}

\end{tabular}
\caption{The projection of the maximum Lorentz factor $\gamma_{\rm max}$ values, in logarithmic color scale, for a sample of $\sim 10^4$ particles along their trajectories onto the PC for $\alpha=45^\circ$ and $\sigma=1\,\Omega$ outside the LC. We see that for the lower conductivity outside the LC the higher maximum $\gamma_{\rm max}$ ($>10^7$) are coming from the trajectories both in the edge and within the interior of the PC. The maximum Lorentz factor can reach up to $\geq 10^8$, and almost come from the particles originating in the edge of the PC. The notations LD, TR, AX, and EQ in the diagrams represent the directions of the leading side edge, trailing side edge, the rotational axis, and the rotational equator in relative to the magnetic axis, respectively.}
\label{fig-pcgam45-1}
\end{figure}

\begin{figure*}
\begin{tabular}{c}
\hspace*{-2cm}
\includegraphics[width=7.8 cm,height=6 cm]{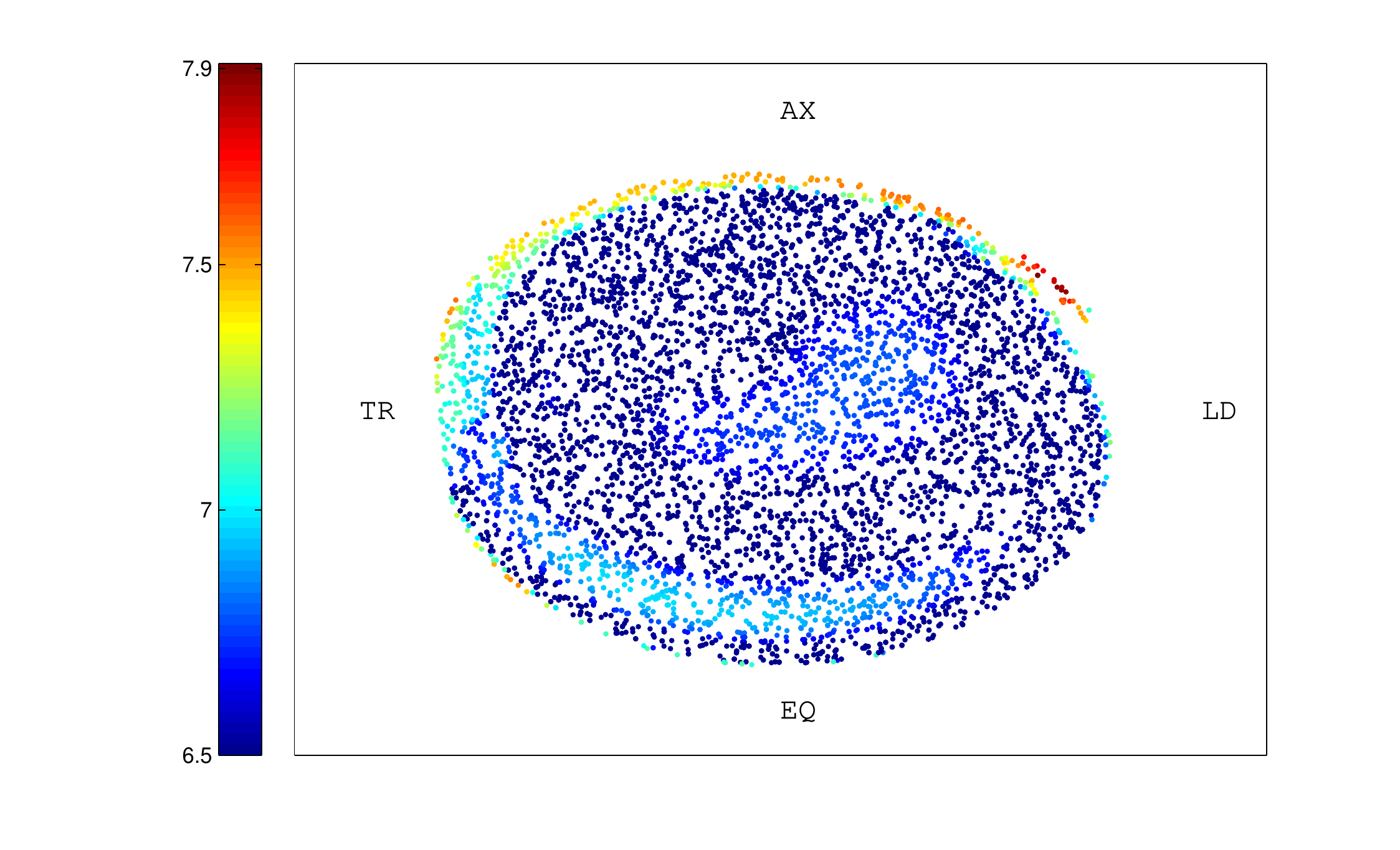}\,
\hspace*{-1cm}
\includegraphics[width=6.5 cm,height=6 cm]{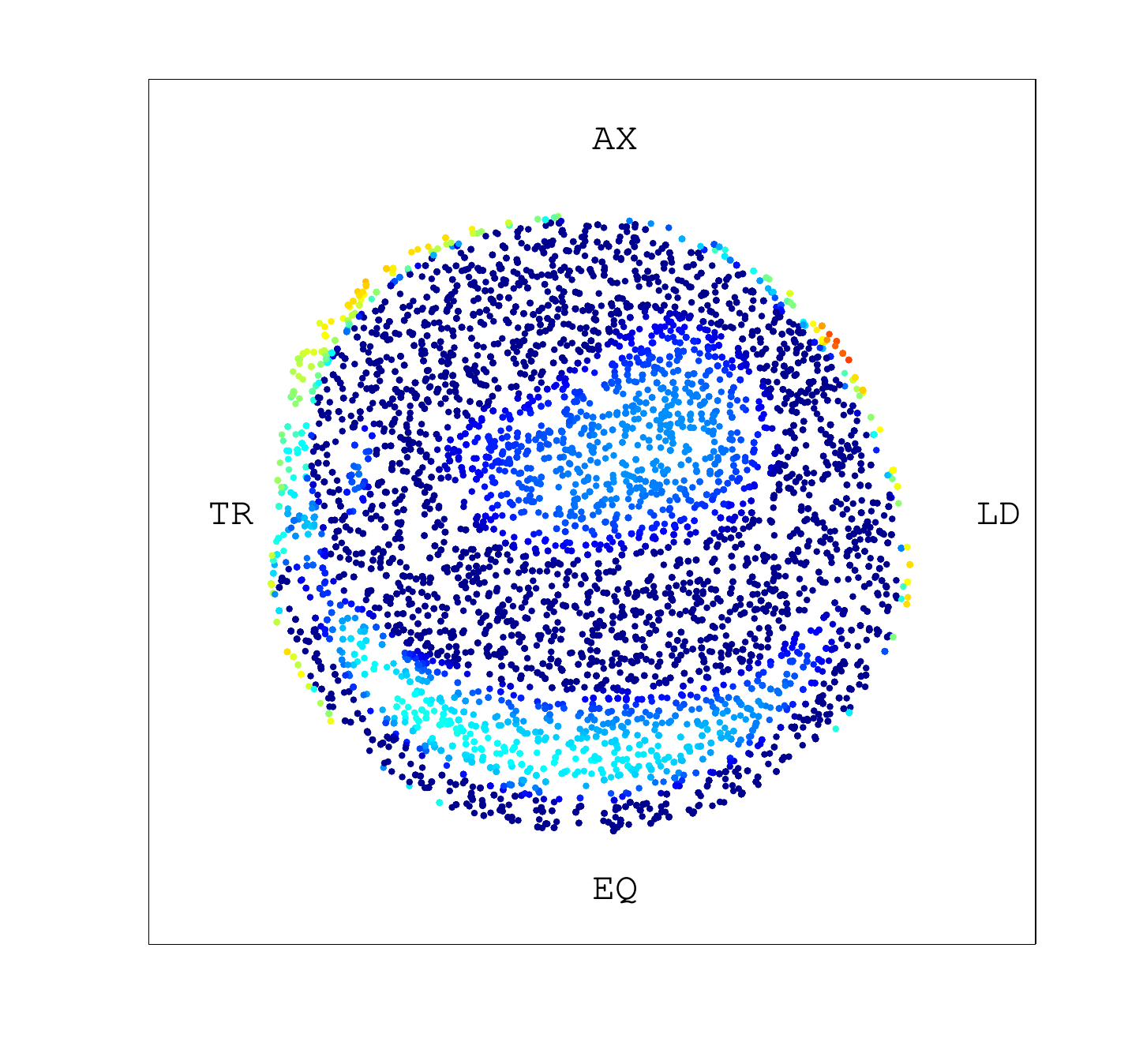}\,
\hspace*{-0.8cm}
\includegraphics[width=6.5 cm,height=6 cm]{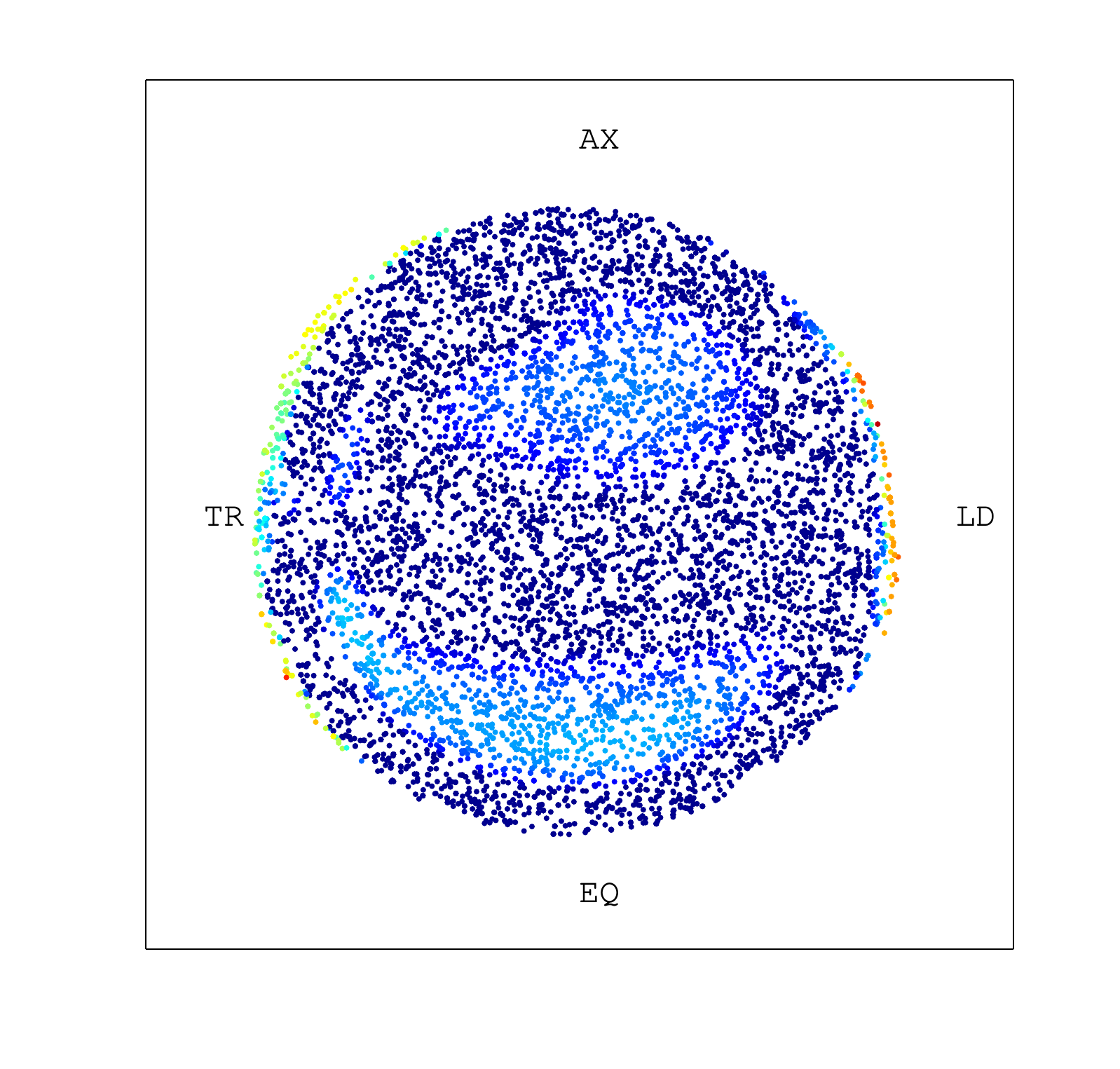}\\
\hspace*{-1.8cm}
\includegraphics[width=7 cm,height=7 cm]{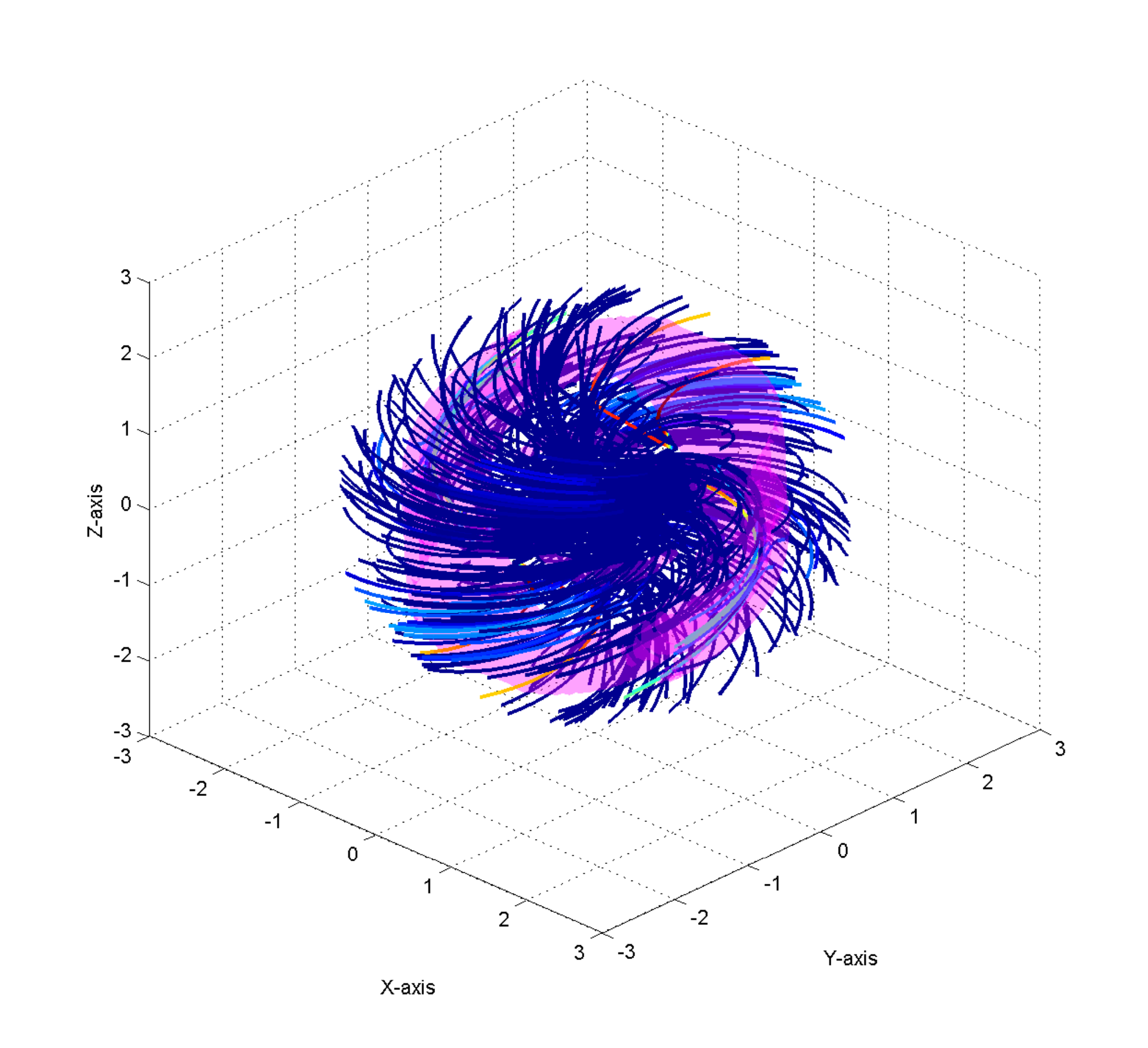}\,
\hspace*{-0.8cm}
\includegraphics[width=7 cm,height=7 cm]{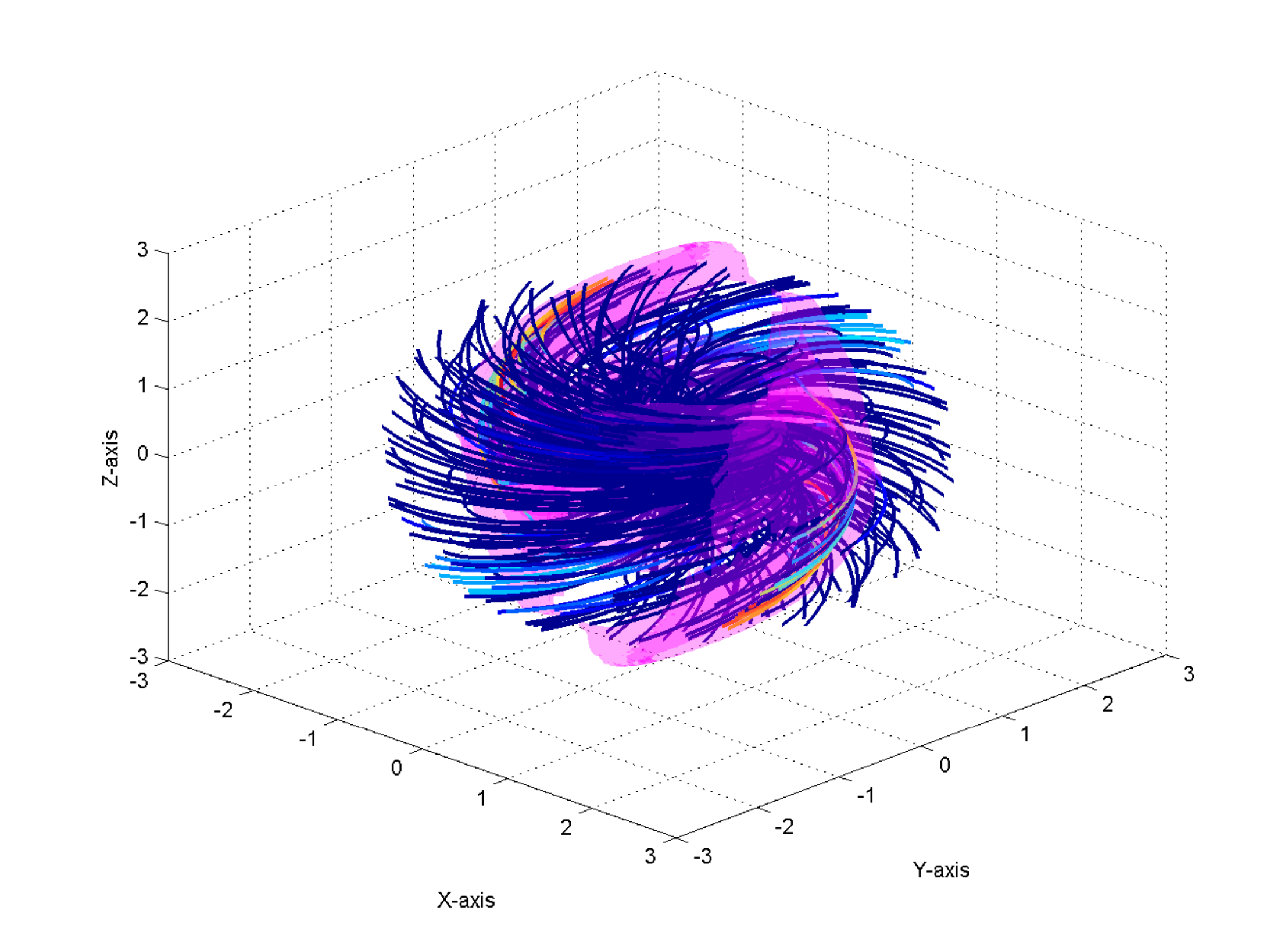}\,
\hspace*{-1.0cm}
\includegraphics[width=7 cm,height=7 cm]{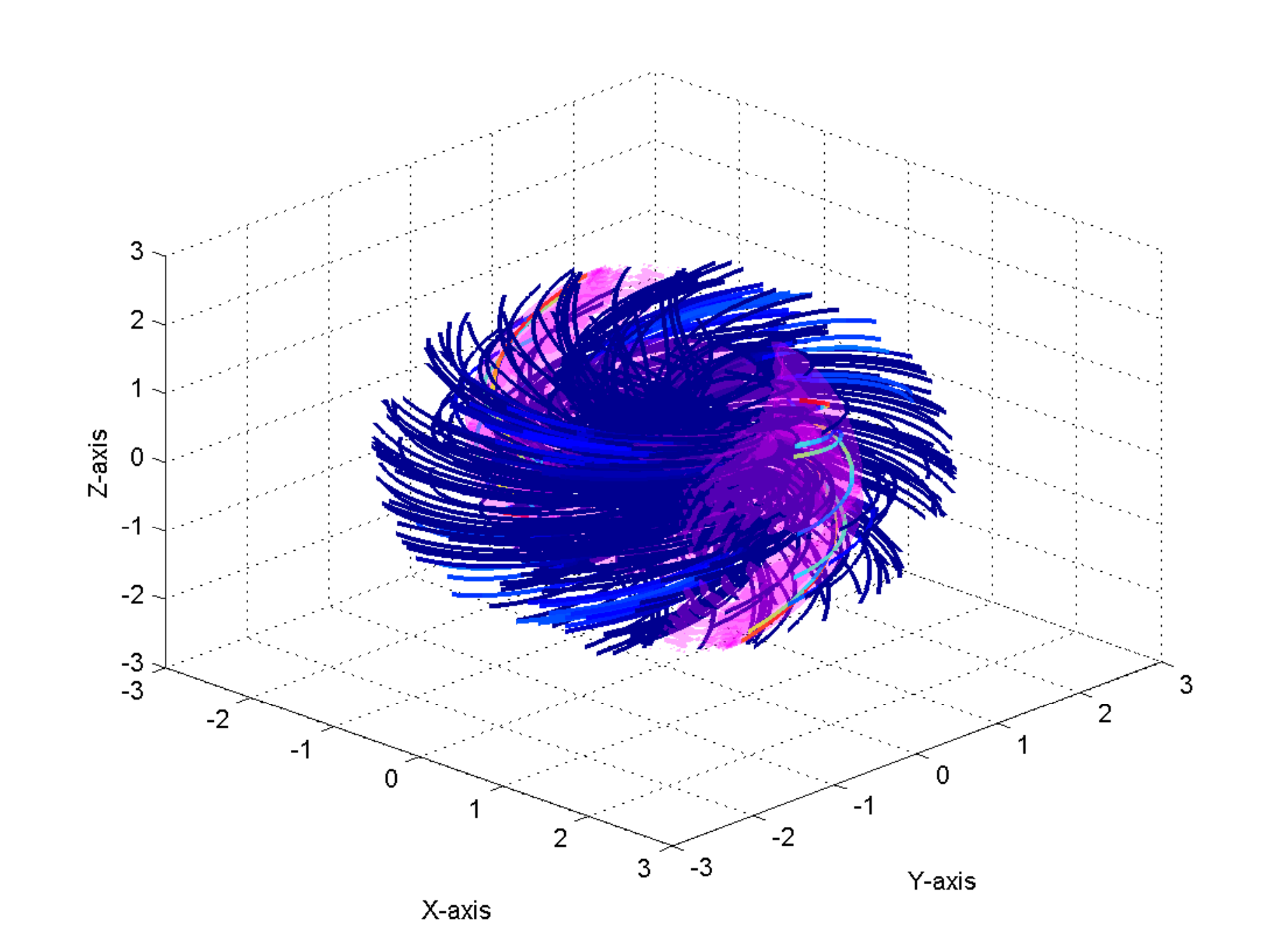}\\
\end{tabular}
\caption{Top row: the same as Fig (\ref{fig-pcgam45-1}) for magnetic inclination (from the left to the right) $45^\circ$,$60^\circ$, and $75^\circ$, separately, for the FIDO magnetospheres with $\sigma= 30\;\Omega$ outside the LC. We find that the maximum $\gamma_{\rm max}$ values are coming from the particle originating from the leading side of the edge. For the fixed $\sigma$, as the inclination increases, we see that the distributions of the higher Lorentz factors are contracted around the edges of PC and will gradually move toward the LD side of the PC. Bottom row: the magenta surface represents the 3D volume rendering of the equatorial current sheet outside the LC; the projections of the Lorentz factor (in the same logarithmic color scale as the up row) along their trajectories, in the corotating frame, for a sample of 300 particles. We see that the larger Lorentz factors are produced by the trajectories originating from the PC edges and reaching around the equatorial current sheet. Moreover, the standard pulsar parameters $P=0.1 \, \rm s$ and $B_{\star}=10^{12} \, \rm G$ are used.}
\label{fig-pcgam}
\end{figure*}

\begin{figure*}
\center
\begin{tabular}{cccccccccc}
\includegraphics[width=4.5 cm,height=4 cm]{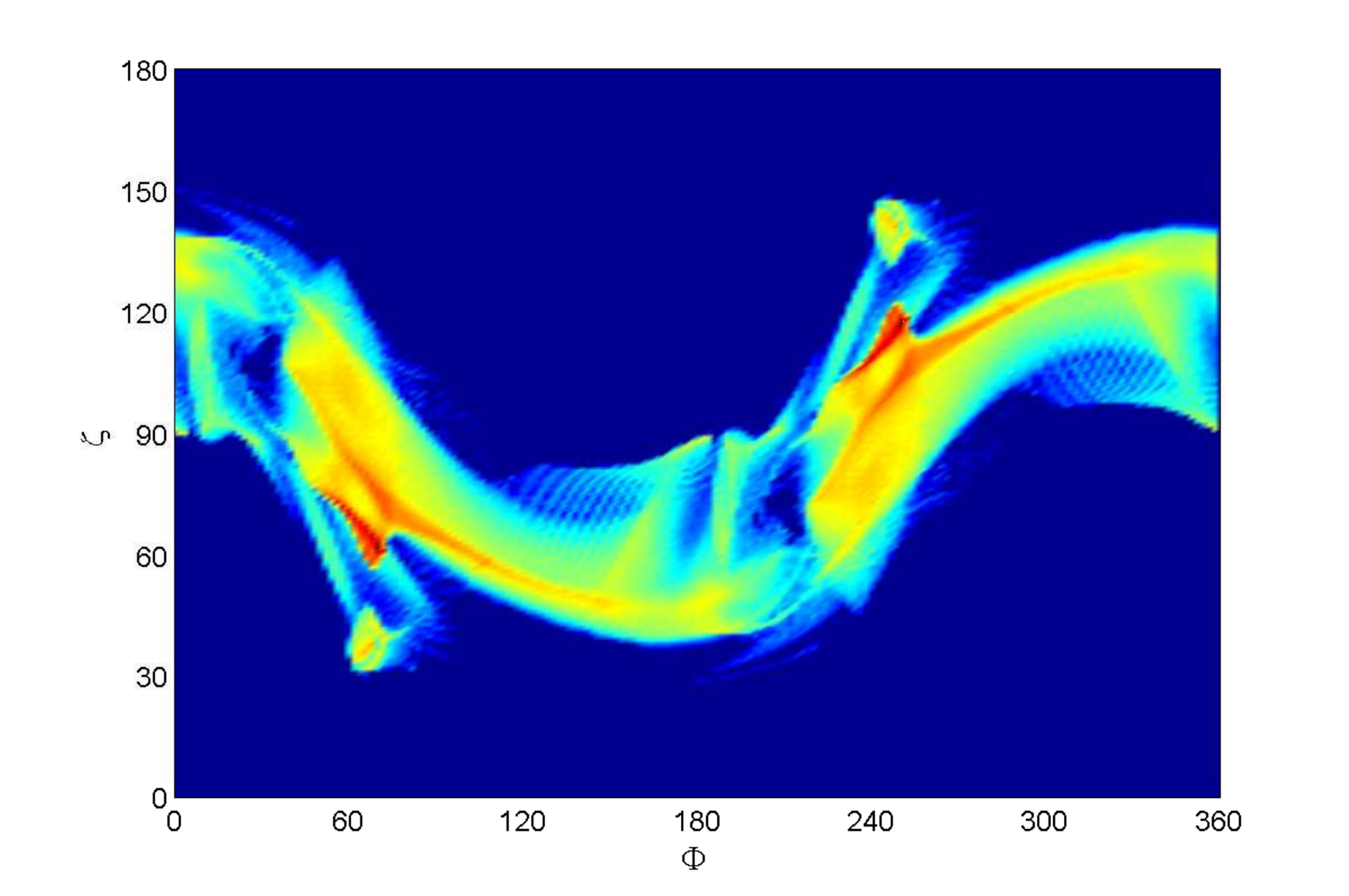}\,
\includegraphics[width=13.5 cm,height=4 cm]{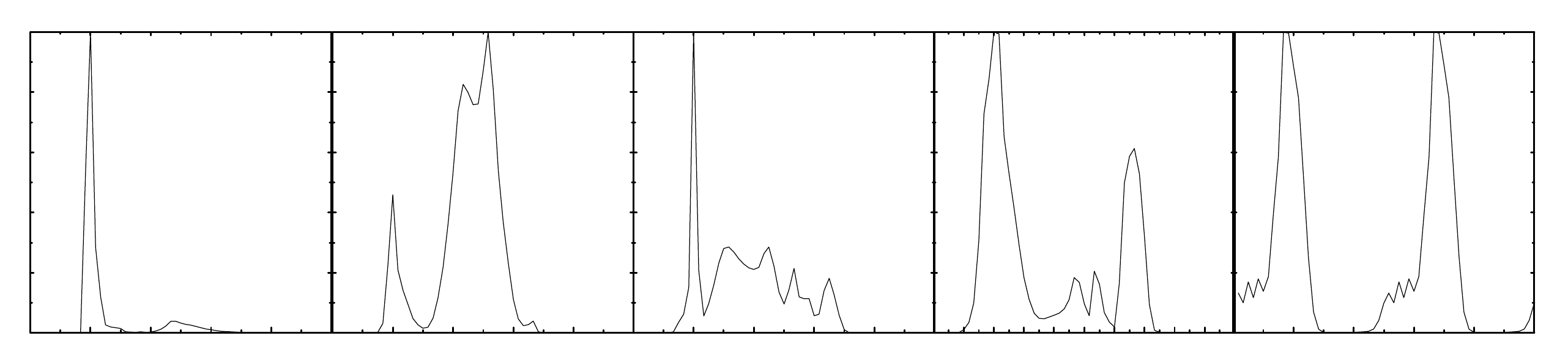}\\
\includegraphics[width=4.5 cm,height=4 cm]{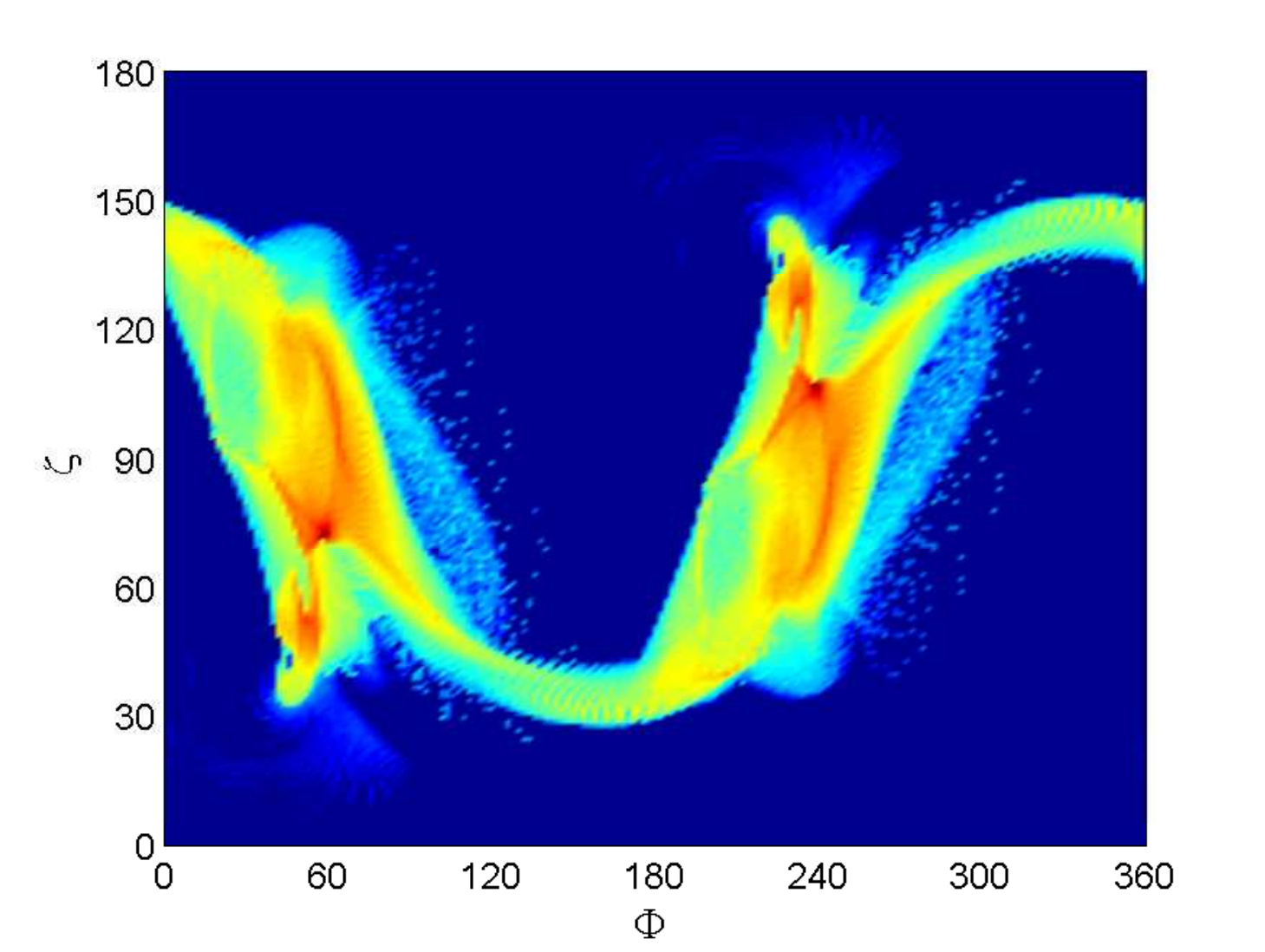}\,
\includegraphics[width=13.5 cm,height=4 cm]{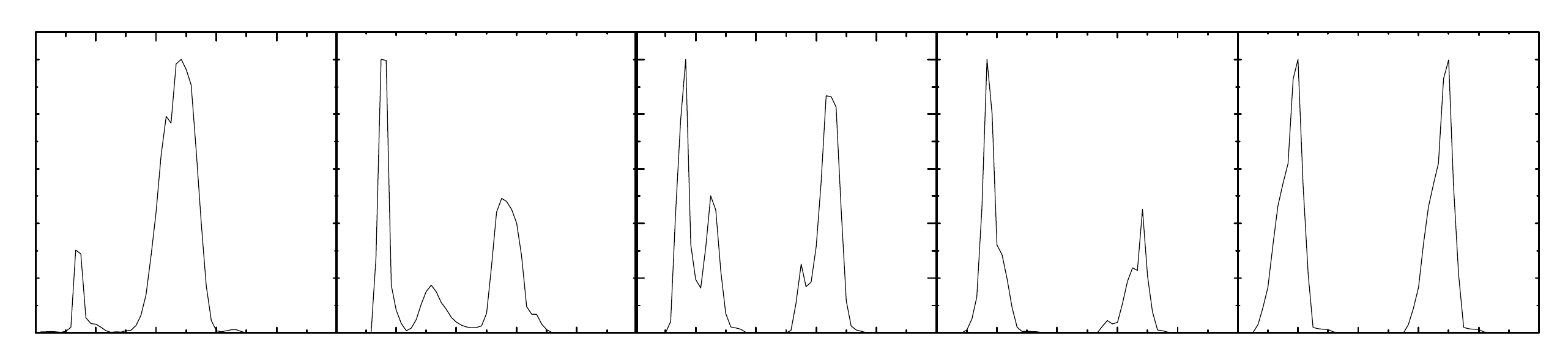}\\
\includegraphics[width=4.5 cm,height=4 cm]{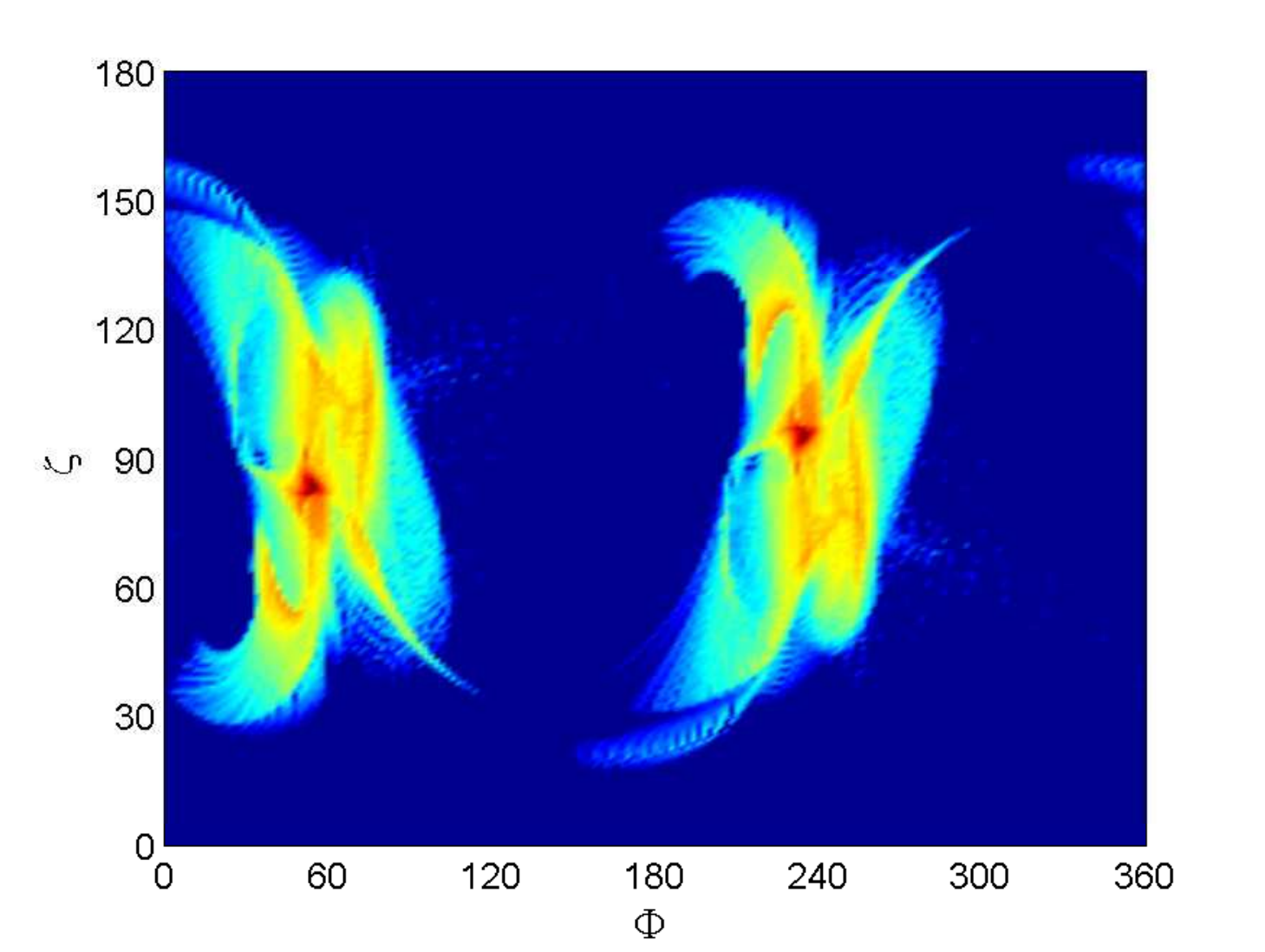}\,
\includegraphics[width=13.5 cm,height=4 cm]{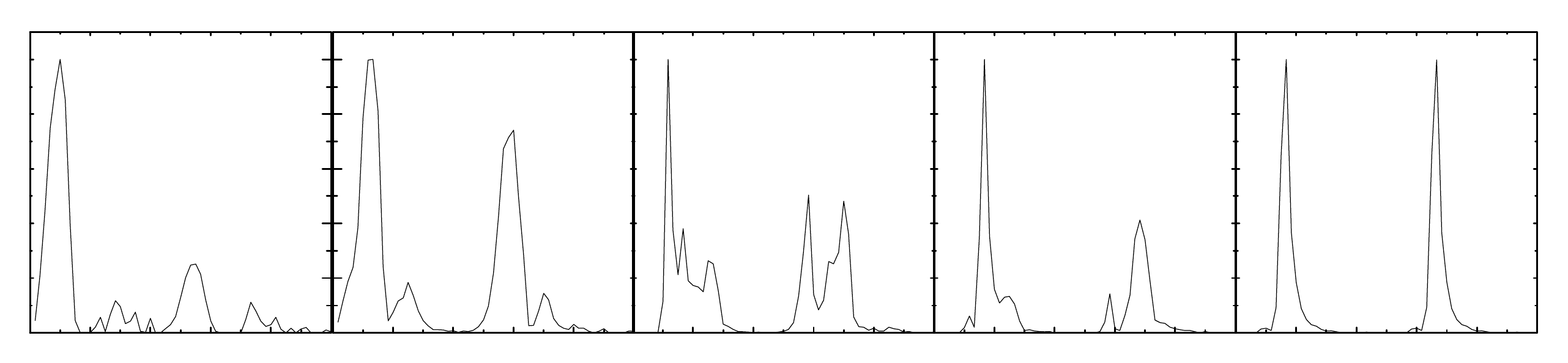}\\
\includegraphics[width=4.5 cm,height=4 cm]{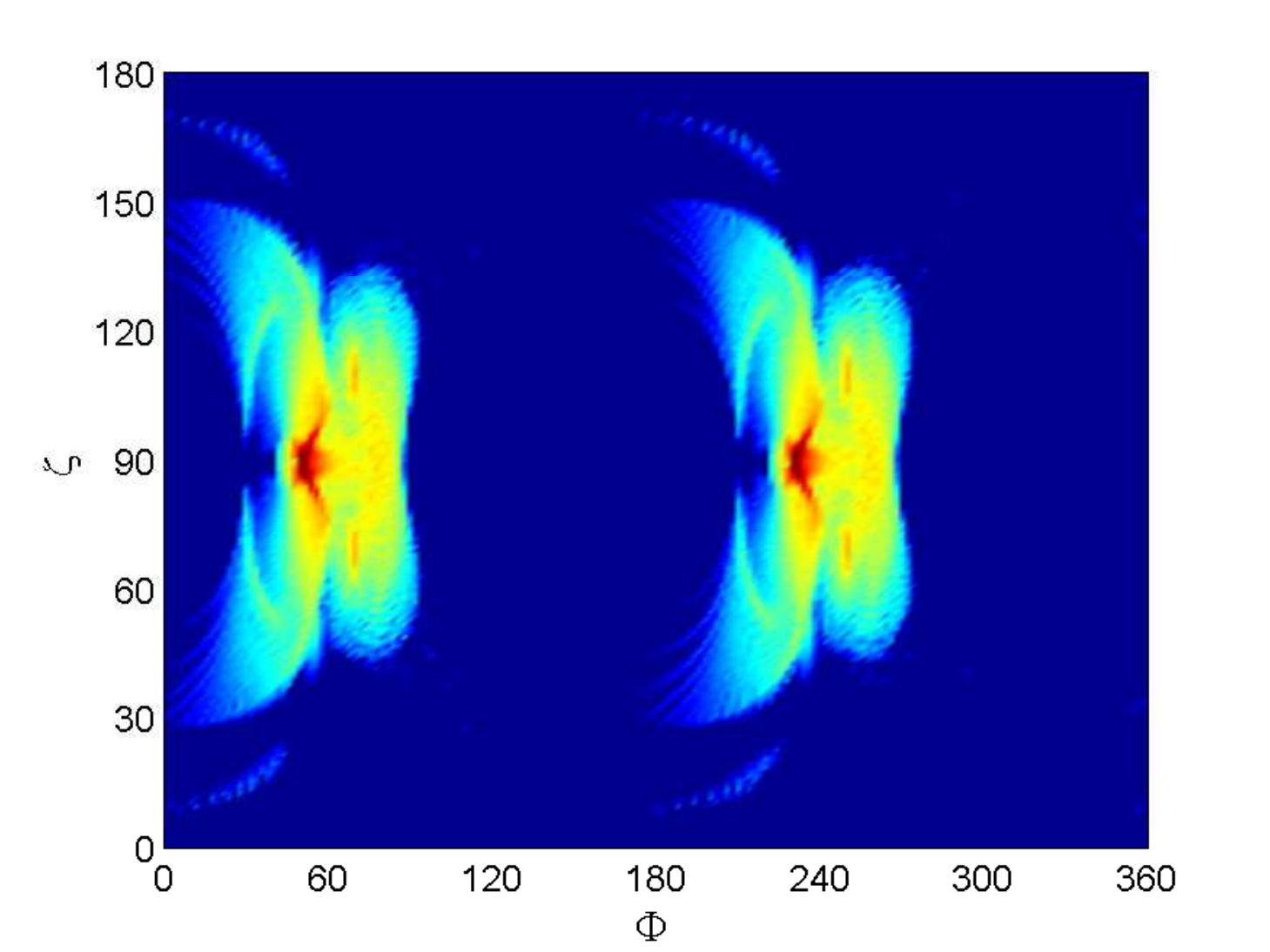}\,
\includegraphics[width=13.5 cm,height=4 cm]{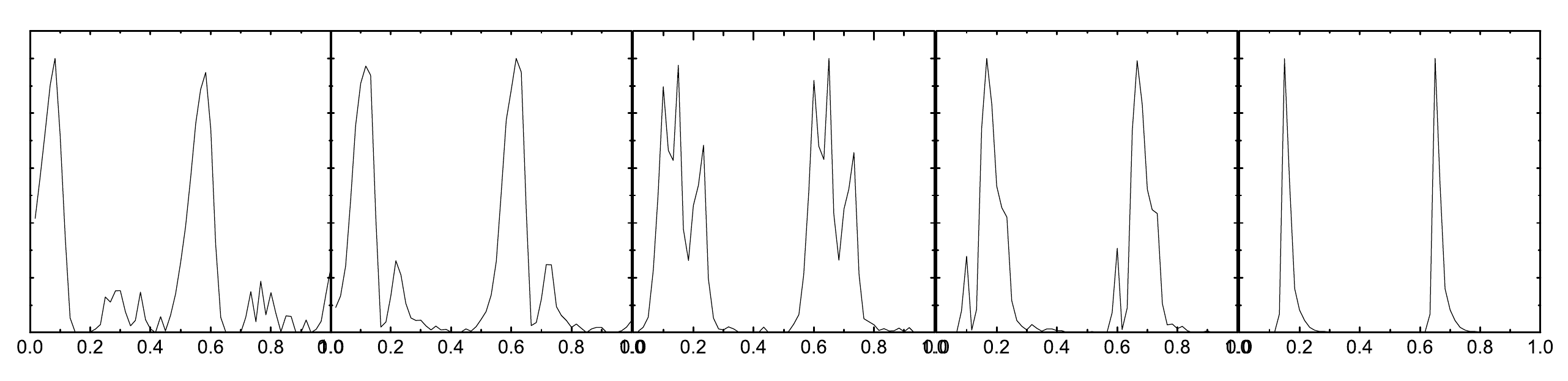}\\
\end{tabular}
\caption{The sky maps and light curves for FIDO magnetospheres with $\sigma=30\;\Omega$ outside the LC for a series of magnetic inclinations $\alpha$=$45^\circ$, $60^\circ$, $75^\circ$, and $90^\circ$ (from the top to the below). For the light curves in each row are obtained by cutting $\zeta=$ $30^\circ$, $45^\circ$, $60^\circ$, $75^\circ$, and $90^\circ$ (from the left to the right). The pulsar parameters $P=0.1 \, \rm s$ and $B_{\star}=10^{12} \, \rm G$ are used.  }
\label{fig-skymap}
\end{figure*}
In Figure (\ref{fig-pcgam45-1}), we plot the maximum Lorentz factor$\gamma_{\rm max}$ values of the particles along the trajectories onto the PC for $\alpha=45^\circ$ and $\sigma=1\,\Omega$, we find that larger Lorentz factors ($\geq) 10^7$ come from the trajectories both in the edge and within the interior of the PC, and that the maximum Lorentz factors reaching up to $(\geq) 10^8$ are almost coming from the particles originating in the edge of the PC. This is due to the fact that, for much lower conductivity values, especially for $\alpha<45^\circ$ and $\sigma<1~\Omega$, the FIDO magnetospheres will significantly deviate from the FF geometry, the equatorial current sheet and the high accelerating electric field will also be destroyed. The high accelerating electric field will distribute in a wider space outside the light cylinder. Similar phenomena were also noticed by other groups \citep{kal14,bra15,cao19}. Therefore, the particles both originating from the edge and the interior of the PC will encounter the high accelerating electric fields and be accelerated to larger Lorentz factors to radiate GeV photons.

Moreover, we note that, as the conductivity increases, the distribution of accelerating electric fields will gradually contract and concentrate mainly near the equatorial current sheet outside the LC, the particles contributing to the GeV emission will almost the ones starting from the edges of the PCs. We also note that the lower (larger) the conductivity values outside the LC, the larger (lower) the lorentz factors. However for much lower conductivity values ($\ll30\,\Omega$) the Lorentz factor $\gamma$ can be accelerated much easier up to $\geq 10^{8}$. For the larger conductivity values ($\gg30\,\Omega$)the larger Lorentz factors will be expected to scarcely exceed $10^{6}$ and this will lead to the MeV emission. These results are due to the fact that the larger conductivity values always restrain the extension and strength of the accelerating electric fields in the equatorial current sheet. Therefore, the appropriate values of the conductivity used to generate the light curves and GeV photons consistent with the ones observed are important in the FIDO models. Similar conclusions are also obtained by \citet{kal14,cao19}.

In the top row of Figure (\ref{fig-pcgam}), we project the maximum values ($\gamma_{\rm max}$) of the particles along each trajectory onto the PCs of the pulsars. We see that the maximum $\gamma_{\rm max}$ values almost comes from the trajectories originating from the leading side of the PC edges, which are the origin of the dissipative regions outside the LC, where almost all the effective radiation particles are coming from. Because particles originating from the PC edges will reach the equatorial current sheet around the rotating equator, encounter the high $E_{||}$ values , and are quickly accelerated to the $\gamma_{\rm max}$ values. We also see that the distributions of $\gamma_{\rm max}$ are asymmetric around the PC edges and all the  $\gamma_{\rm max}$ values are almost concentrated on the leading side of PC edges. The asymmetries of the high $\gamma_{\rm max}$ values around the PC also indicate that the effects of the equatorial current sheet on the particles are not symmetrical. Moreover, we also notes that distributions of the high $\gamma_{\rm max}$ values shrink with the increasing $\alpha$ for the fixed $\sigma$ due to the decreasing dissipative region in the current sheets \citep{kal14,cao20}.

In bottom row of Figure (\ref{fig-pcgam}),  we plot the 3D volume rendering of the equatorial current sheet outside the LC and the projections of the Lorentz factor along their trajectories, in the corotating frame, for a sample of 300 particles. We note that the larger Lorentz factors are indeed produced by the trajectories originating from the PC edges and reaching around the equatorial current sheet outside the LC.

\begin{figure*}
\center
\begin{tabular}{c}
\includegraphics[width=3.6 cm,height=3.6 cm]{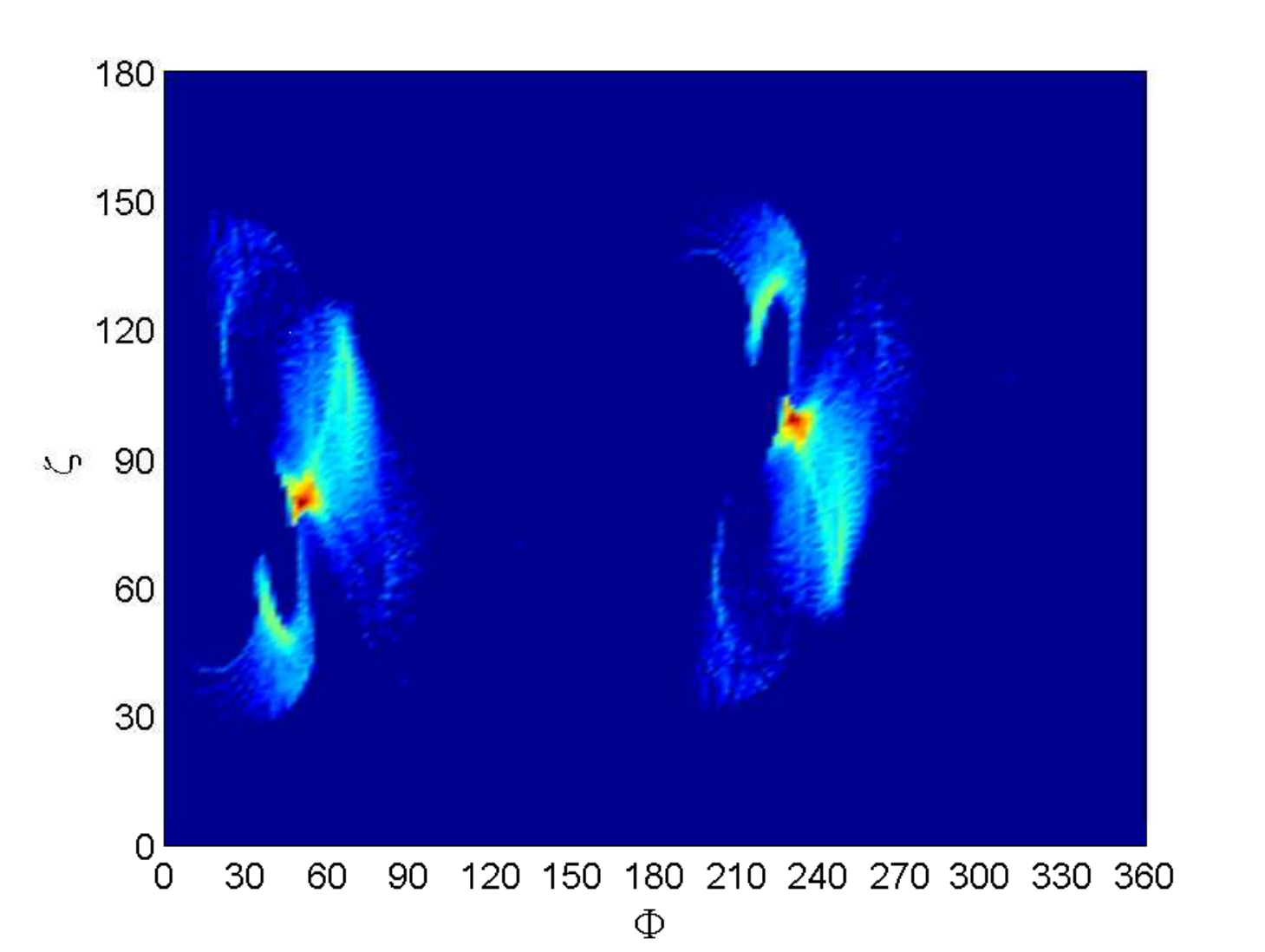}
\includegraphics[width=3.6 cm,height=3.6 cm]{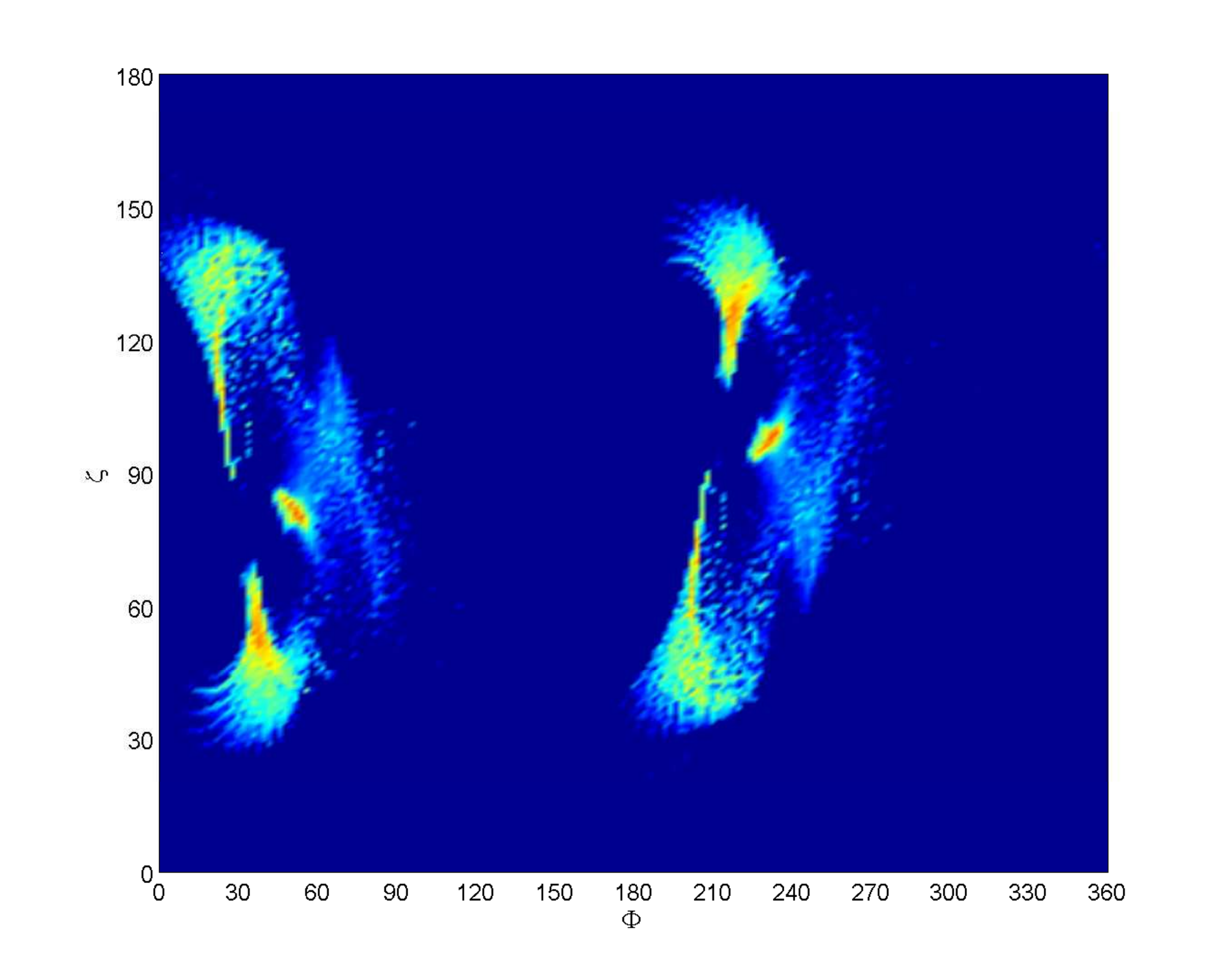}
\includegraphics[width=3.6 cm,height=3.6 cm]{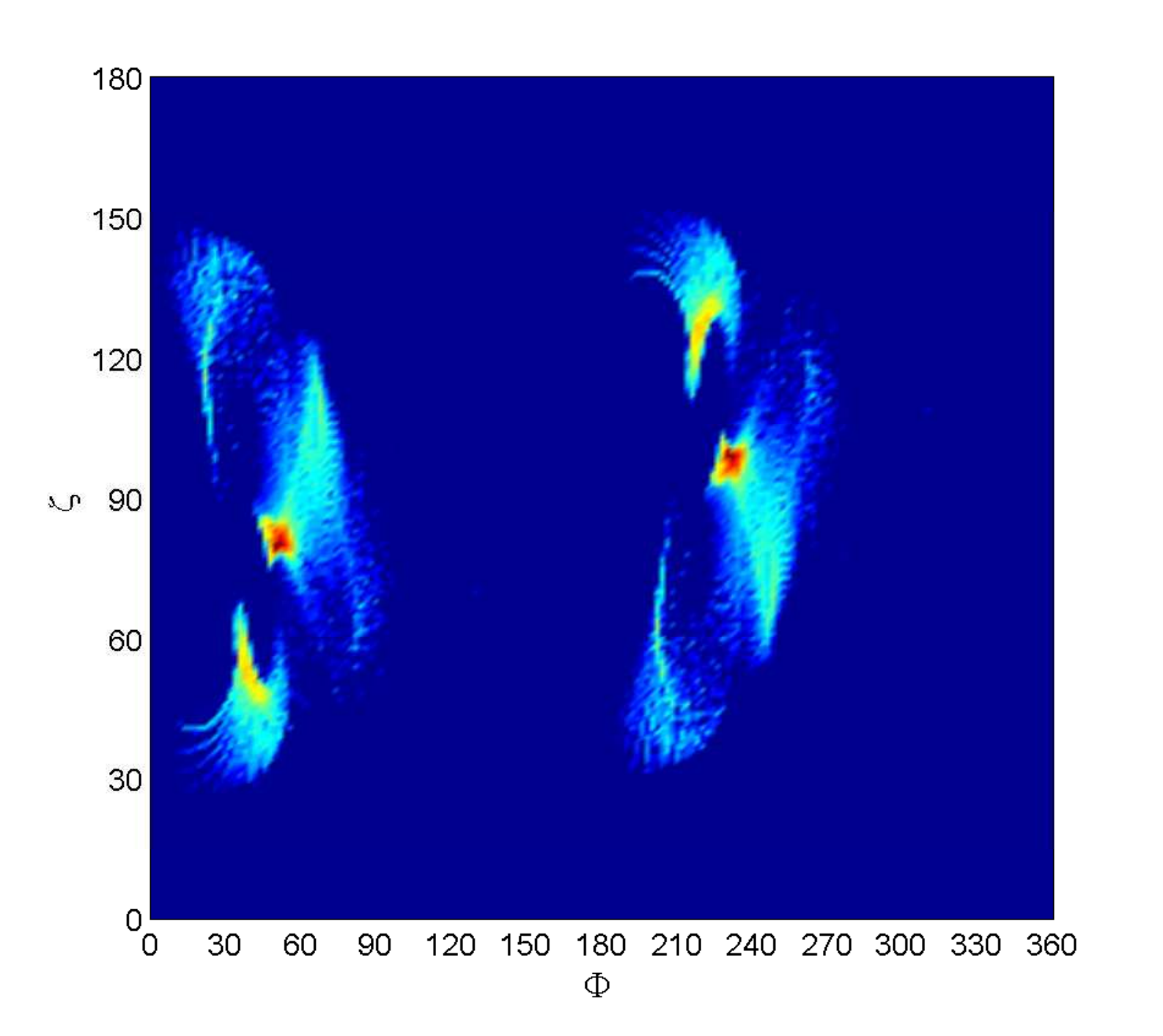}
\includegraphics[width=3.6 cm,height=3.6 cm]{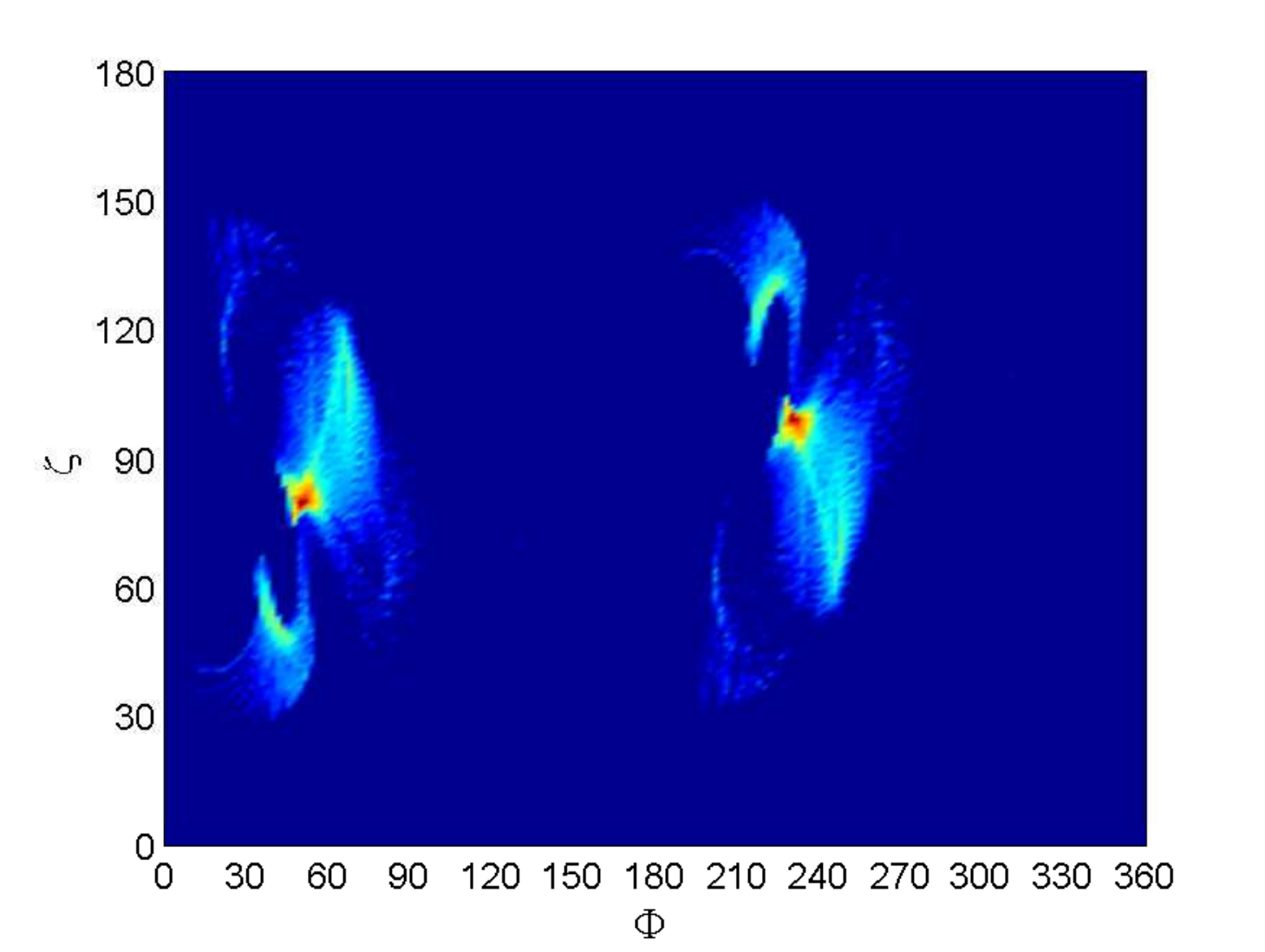}
\includegraphics[width=3.6 cm,height=3.6 cm]{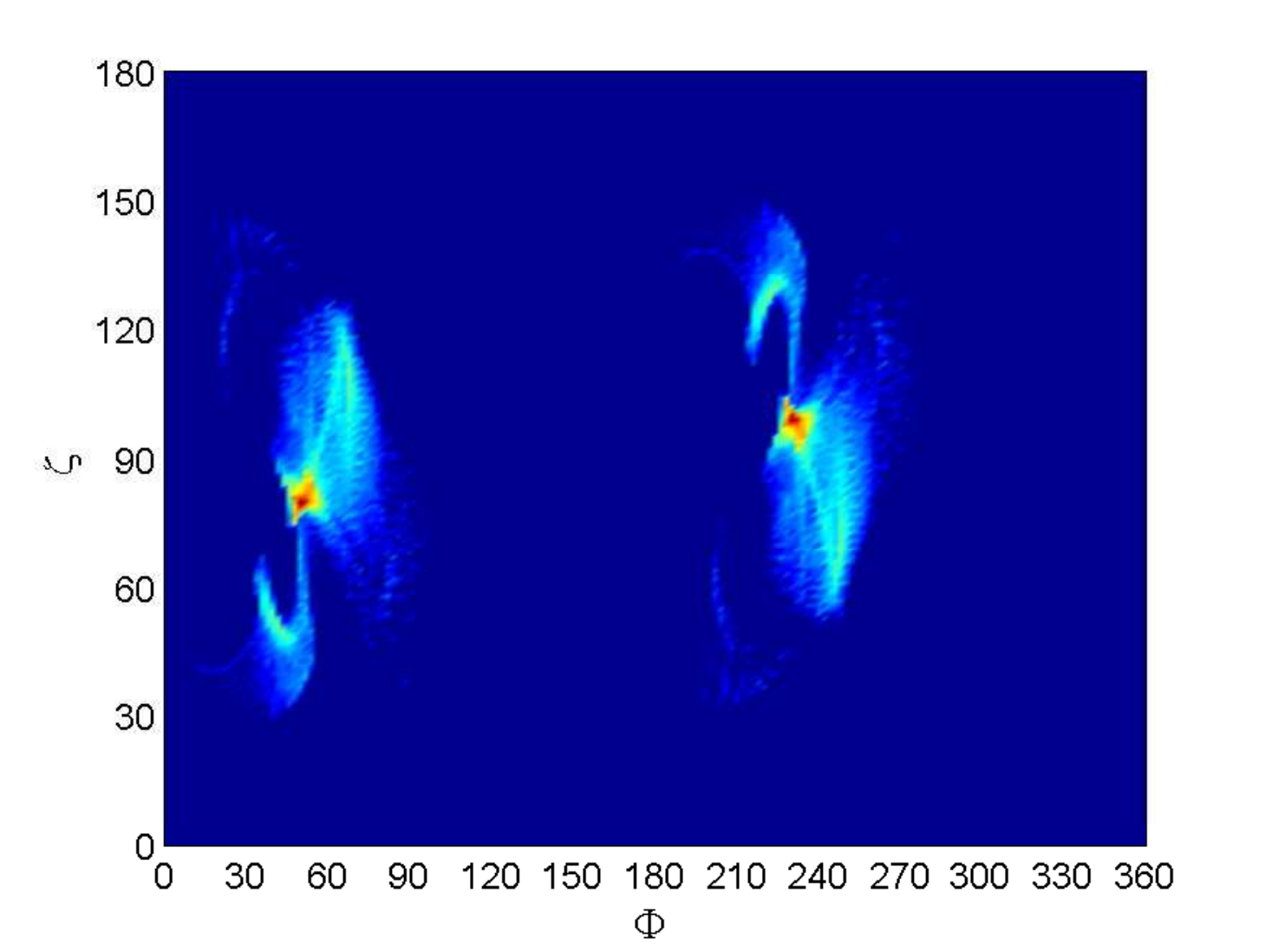}\\
\includegraphics[width=18 cm,height=3.6 cm]{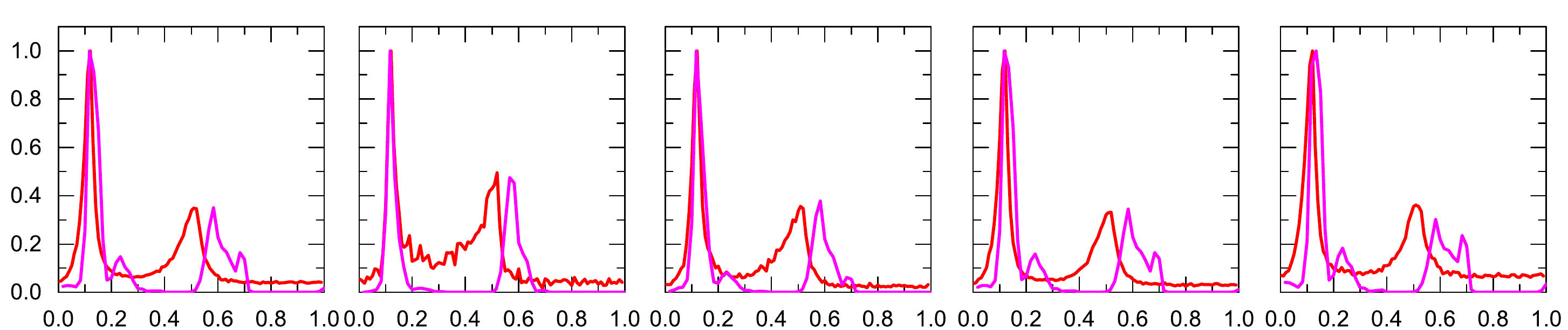}\\

\end{tabular}
\caption{The evolution of energy-dependent sky maps and the light curves of Crab pulsar for $\alpha=75^\circ$, and $\sigma=10\;\Omega$ outside the LC. Top panels are the sky maps for various energy bands of (0.1\,GeV, 50\,GeV), (3\,GeV, 50\,GeV), (1\,GeV,3\,GeV), (0.3\,GeV,1\,GeV), and (0.1\,GeV,0.3\,GeV), from the left to the right, respectively. Bottom panels: corresponding modeling light curves (magenta) are obtained by cutting the sky maps in $\zeta=50^\circ$ and the observed light curves (red) taken from 2PC \citep{abd13}.}
\label{fig-Crab}
\end{figure*}

\begin{figure*}
\center
\begin{tabular}{ccccccccc}
\includegraphics[width=3.6 cm,height=3.6 cm]{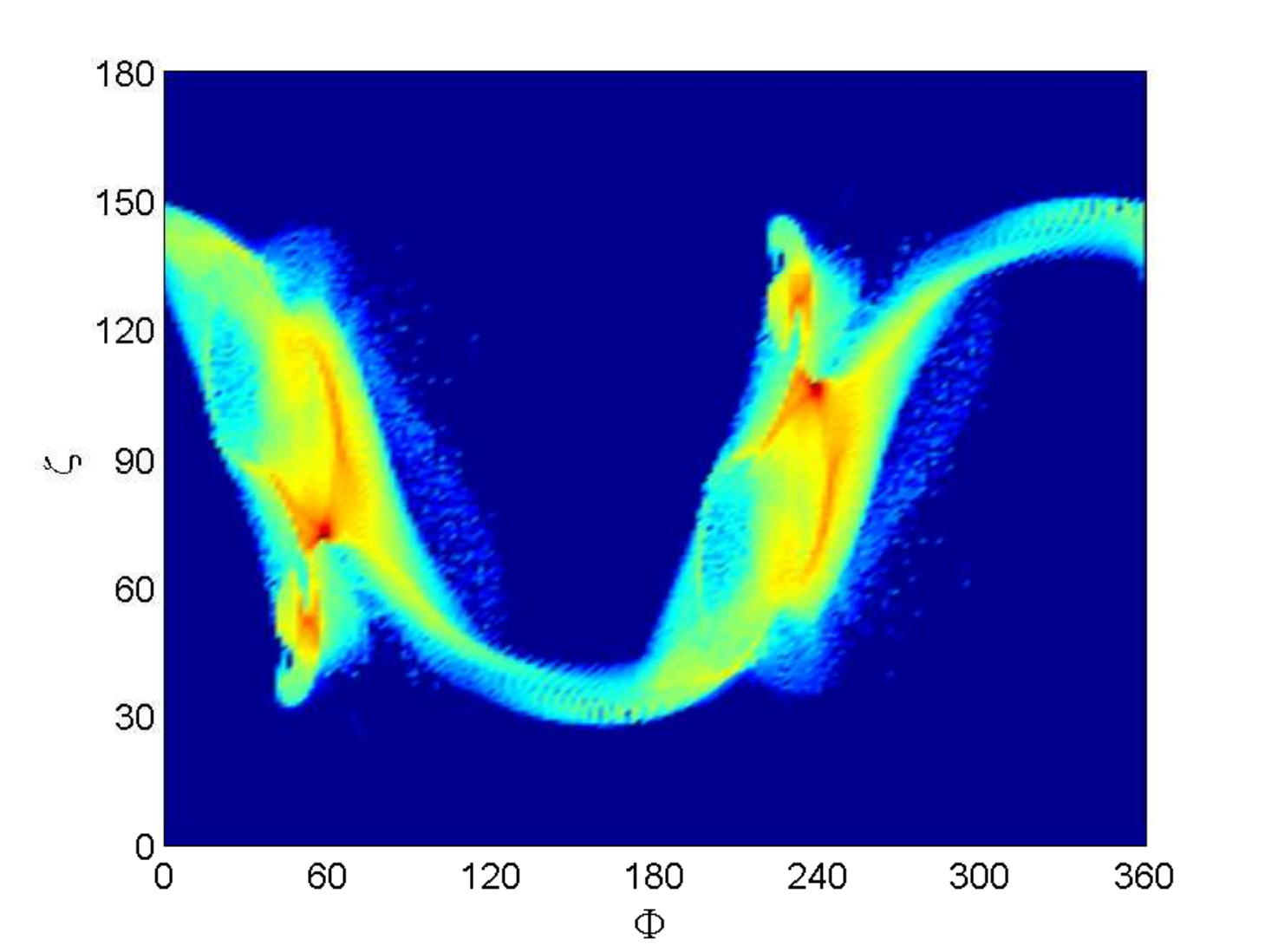}
\includegraphics[width=3.6 cm,height=3.6 cm]{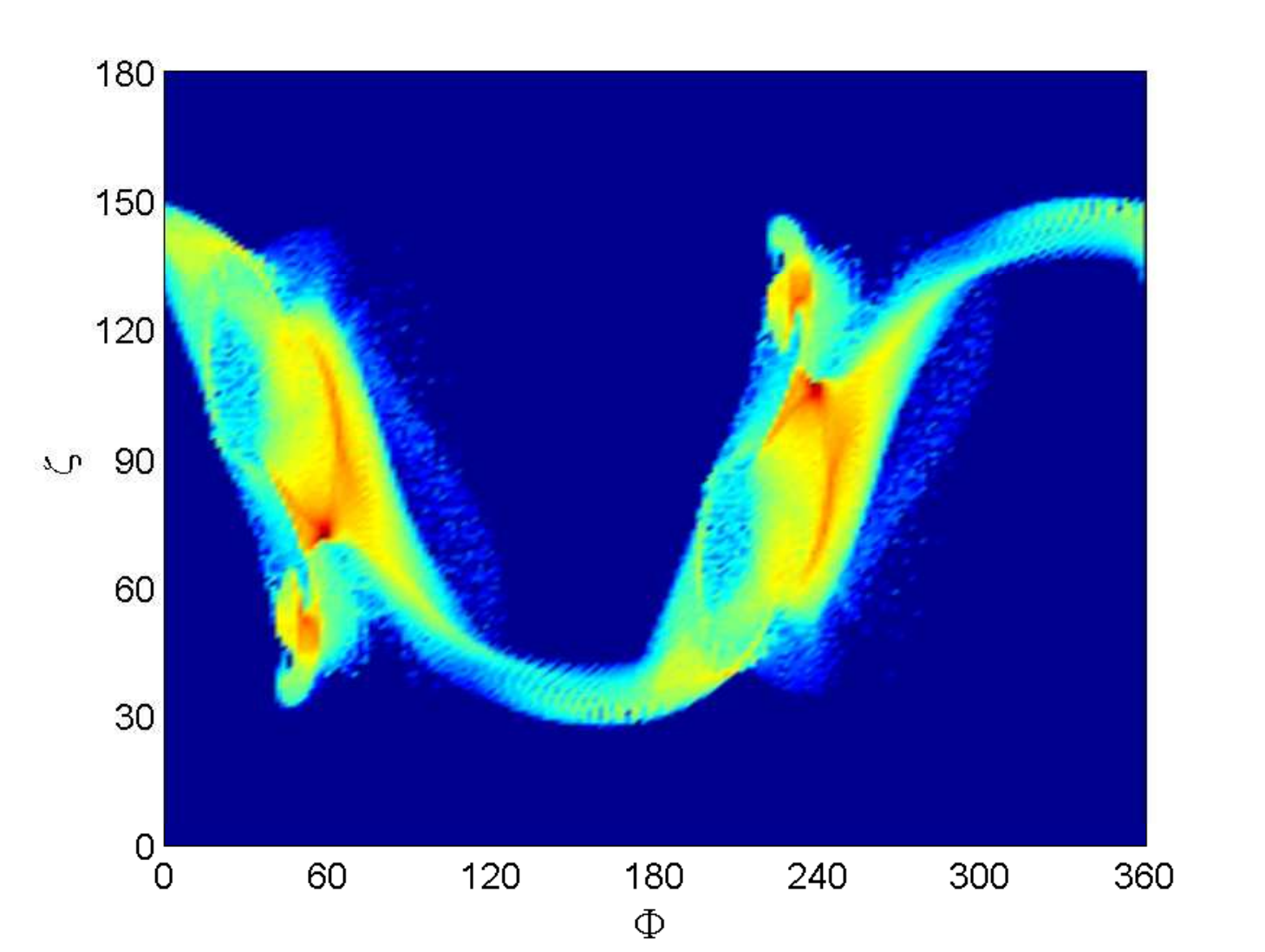}
\includegraphics[width=3.6 cm,height=3.6 cm]{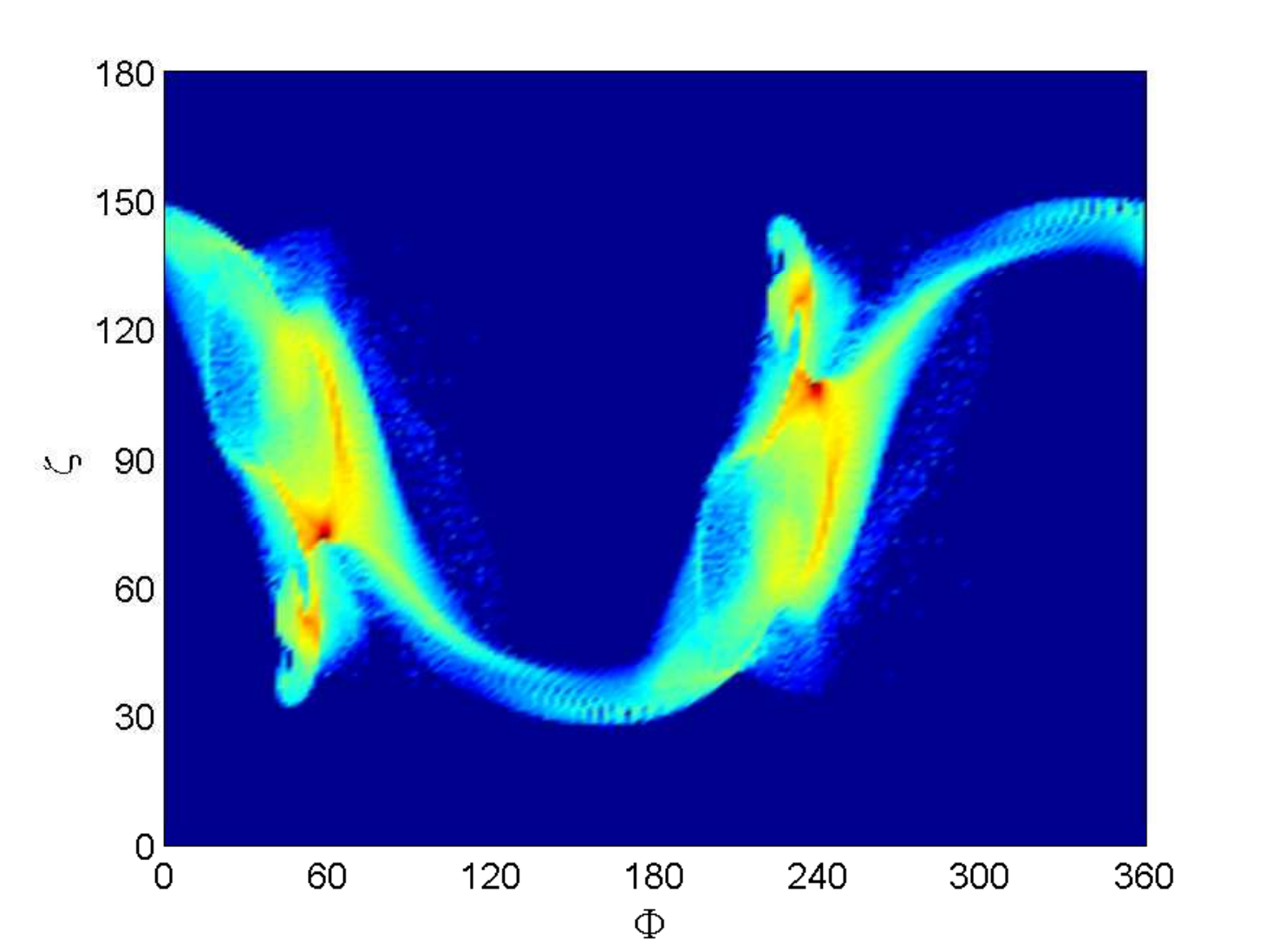}
\includegraphics[width=3.6 cm,height=3.6 cm]{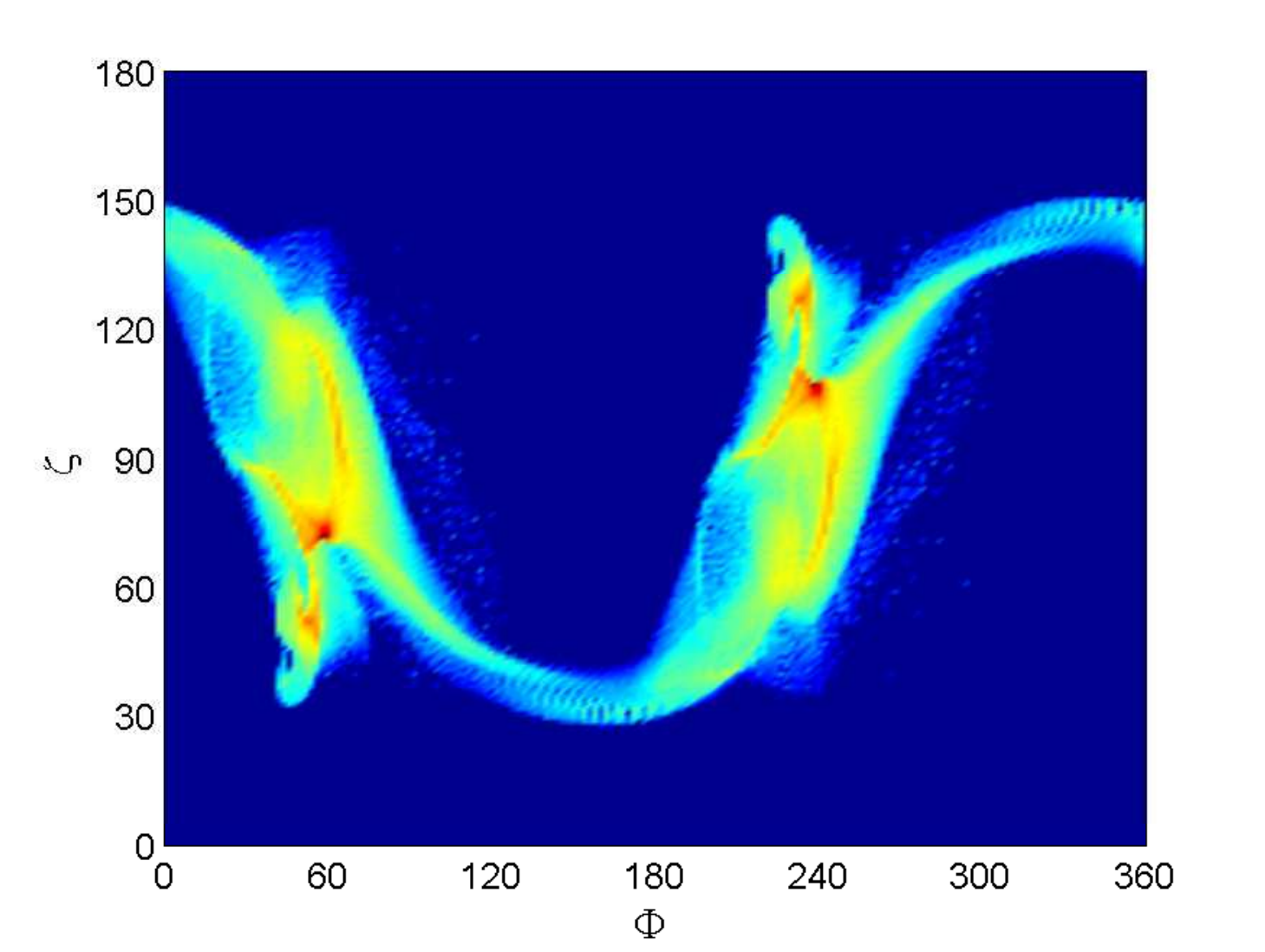}
\includegraphics[width=3.6 cm,height=3.6 cm]{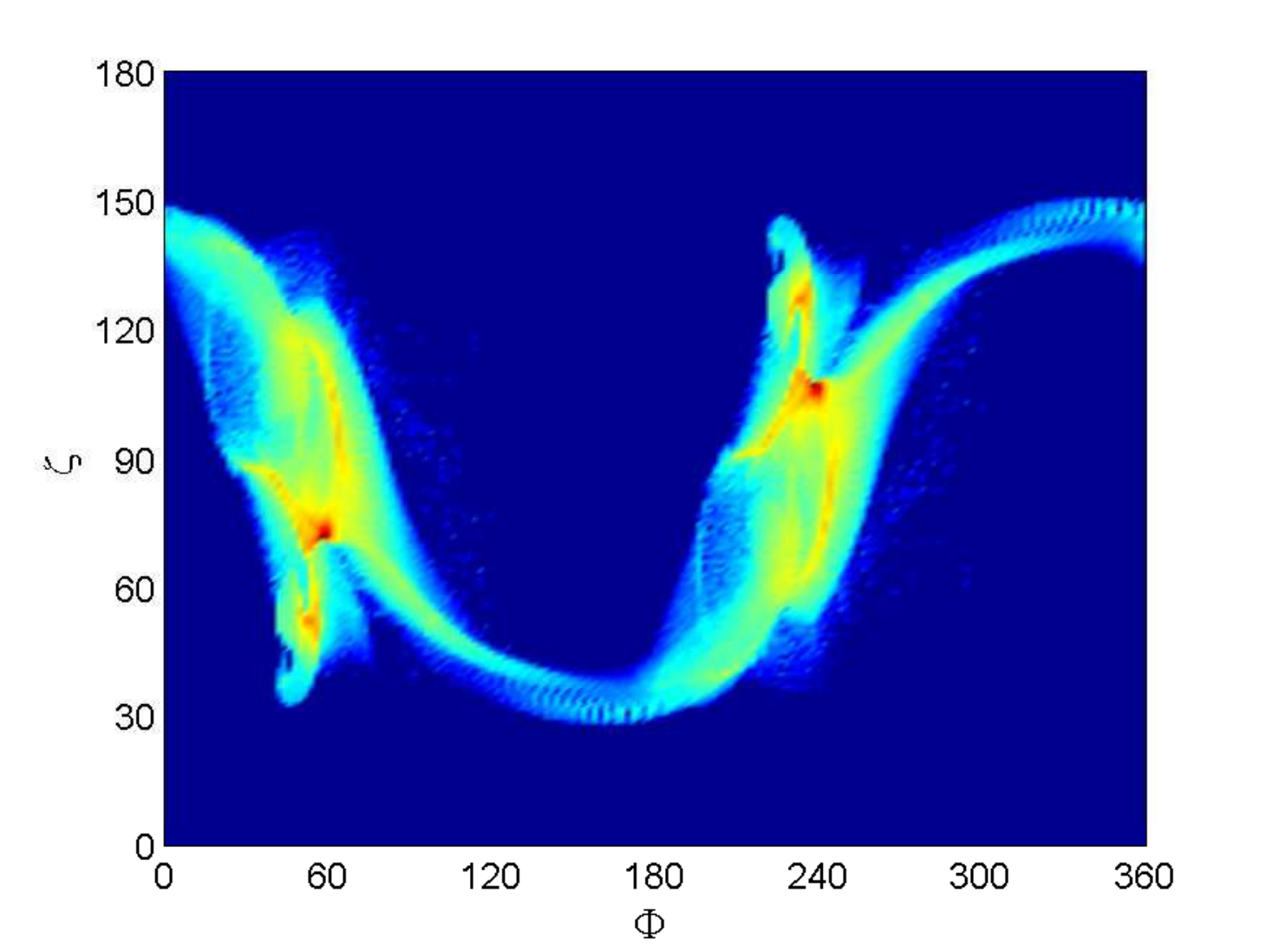}\\
\includegraphics[width=18 cm,height=3.6 cm]{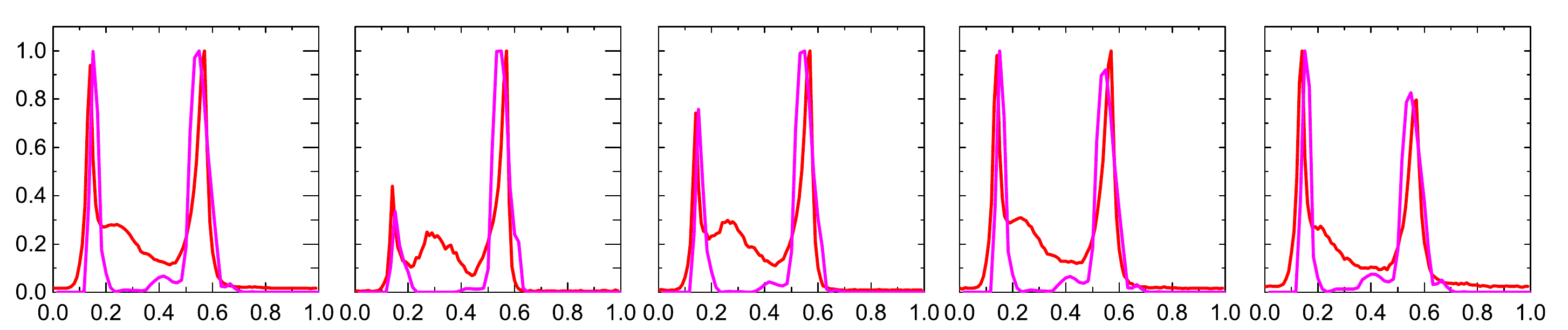}\\
\end{tabular}
\caption{Same as the panels in Figure (\ref{fig-Crab}), but for Vela pulsar $\alpha=60^\circ$, $\sigma=30\;\Omega$ ,and $\zeta=42^\circ$. }
\label{fig-Vela}
\end{figure*}

\section{The energy-dependent sky maps and light curves}\label{sect-SM}
The sky maps and light curves is producing by collecting the bolometric luminosity from all the emitting particles in \citet{cao19}. Here, we expand the previous study by computing the curvature radiation spectrum. We construct the sky maps and light curves by collecting the curvature photons from all the  emitting particles. To construct the pulsar light curves, we need to determine emission direction of the photon in the IOF. The emission direction of the photon ${\bm \eta}_{\rm em}$ is assumed to be along the direction of particle motion ${\bm \beta}={\bf v}/c$  in the IOF, the viewing angles and observing (azimuthal) phase are defined by $\zeta$ and $\phi$, respectively, relative to the pulsar rotation axis
\begin{eqnarray}
\zeta=\rm acos(\beta_{z})\;,
\end{eqnarray}
and
\begin{eqnarray}\label{rot-phase}
\phi=\phi_{\rm rot}-\phi_{\rm em}-{\bf r_{\rm em}} \cdot {\bm \eta}_{\rm em}/R_{\rm LC}\;,
\end{eqnarray}
where the effects of the field rotation and time-delay are taken into account. $\phi_{\rm rot}=\Omega\,t$ is the rotation phase, $\phi_{\rm em}=\arctan(\beta_y/\beta_x)$ is the phase of the emitting photon, and the third term is the phase from the time delay correction.

For given energy region (say 0.1 GeV to 50 GeV), the  curvature photons $(N_{\rm ph})$ from all the emitting particles at the  distance $\bf r$ along their trajectories  are collected  by integrating the curvature spectrum in the energy interval $(E_{\gamma1},E_{\gamma2})$,
\begin{eqnarray}\label{Nphr}
N_{\rm ph}({\bf r})=\int_{\rm E_{\gamma1}}^{\rm E_{\gamma2}}{F(\rm E_{\gamma},r)}\;{\rm d{\rm E_{\gamma}}}\;.
\end{eqnarray}
In our calculations, the observed  phase $\phi$ is uniformly divided into 100 bins (between $0^\circ$ and $360^\circ$), 180 bins for the $\zeta$ ( between $0^\circ$ and $180^\circ$ ), and 220 bins for the energy (between 0.1 GeV and 50 GeV). Here a Gauss profile  is used to smooth the collected curvature photons with  $\Delta\zeta =4^{\circ}$ at view angle $\zeta^0$ by
\begin{eqnarray}\label{Nphs}
N_{\rm ph}\propto \exp \left(-\frac{(\zeta-\zeta^0)^2}{2{\Delta\zeta}^2} \right)\;.
\end{eqnarray}

The sky maps are produced by collecting all the curvature photons from each emitting particle, the light curves are then obtained by cutting the sky maps in a fixed viewing angle $\zeta^0$. We find that the sky maps and light curves are very sensitive on the $\sigma$ values. In the lower $\sigma$ value, the light curves is  broad in shapes and have only one peak, which does not match  with most of the pulsars published in Fermi 2PC (also see,\citet{kal14,cao19}). While as the $\sigma$ values increase, the peaks become narrow and more double-peaks light curves will appear.

In figure (\ref{fig-skymap}), we give the sky maps and light curves in energy band $(0.1\,\rm{GeV},50\,\rm{GeV})$ for FIDO magnetosphere with $\sigma=30$ outside the LC for the magnetic inclination angles $\alpha=45^\circ,60^\circ,75^\circ,90^\circ$ and viewing angles $\zeta=30^\circ,45^\circ,60^\circ,75^\circ,90^\circ$. The standard pulsar parameters $P=0.1 \, \rm s$ and $B_{\star}=10^{12} \, \rm G$ are adopted.
We find that the light curves tend to the double-peaked profiles toward increasing $\alpha$ and/or $\zeta$ direction, some of which are similar to the observed ones published in the Fermi 2PC. The light curves are similar to the ones obtained by collecting the bolometric luminosity in larger viewing angles, while different in the lower viewing angles \citep{cao19} for the spectral difference of the phase. Similar conclusions are also found in \citet{kal14} and \citet{bra15}.

\begin{figure*}
\center
\begin{tabular}{c}
\includegraphics[width=5.5 cm,height=4.5 cm]{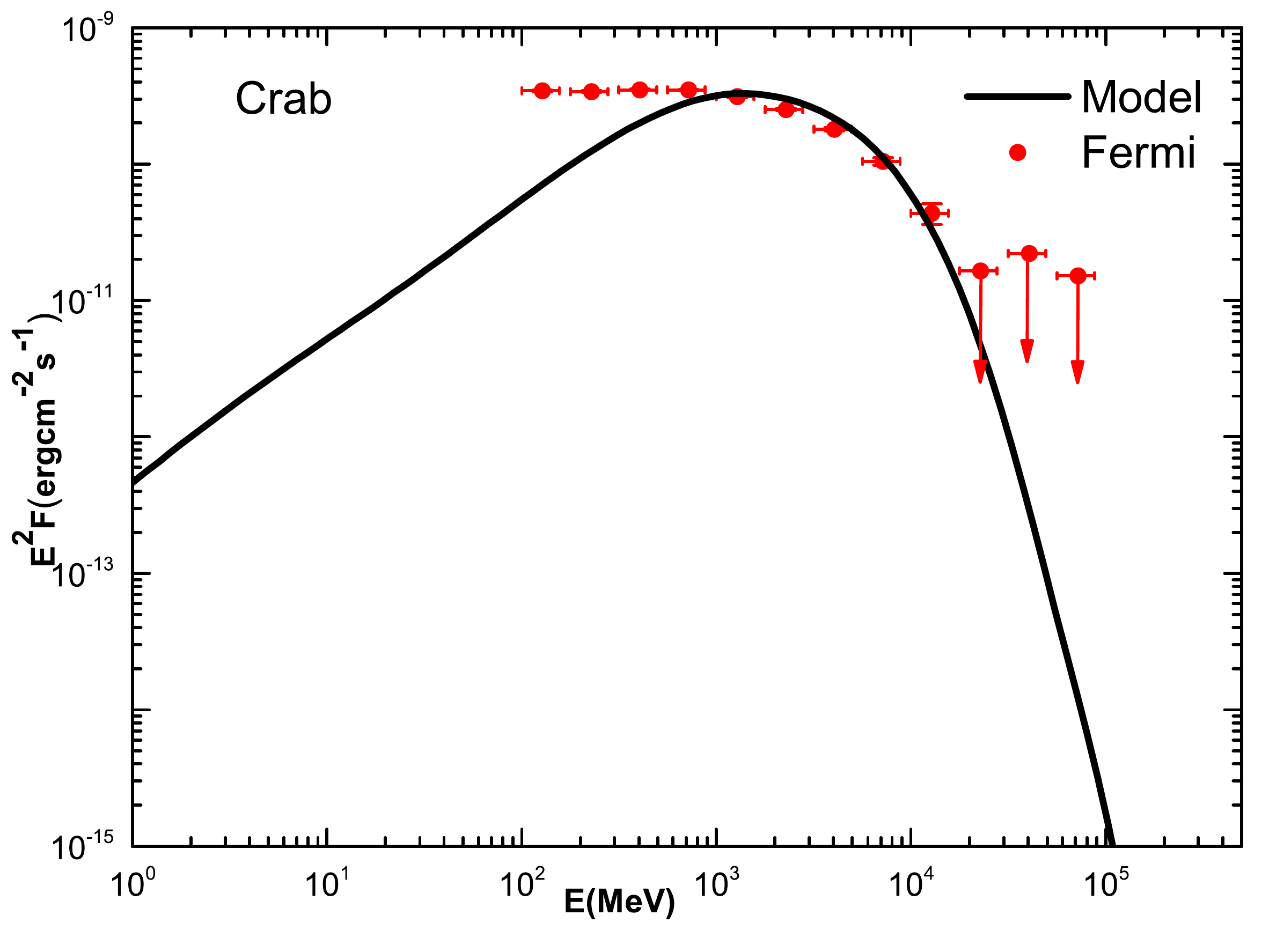}
\includegraphics[width=5.5 cm,height=4.5 cm]{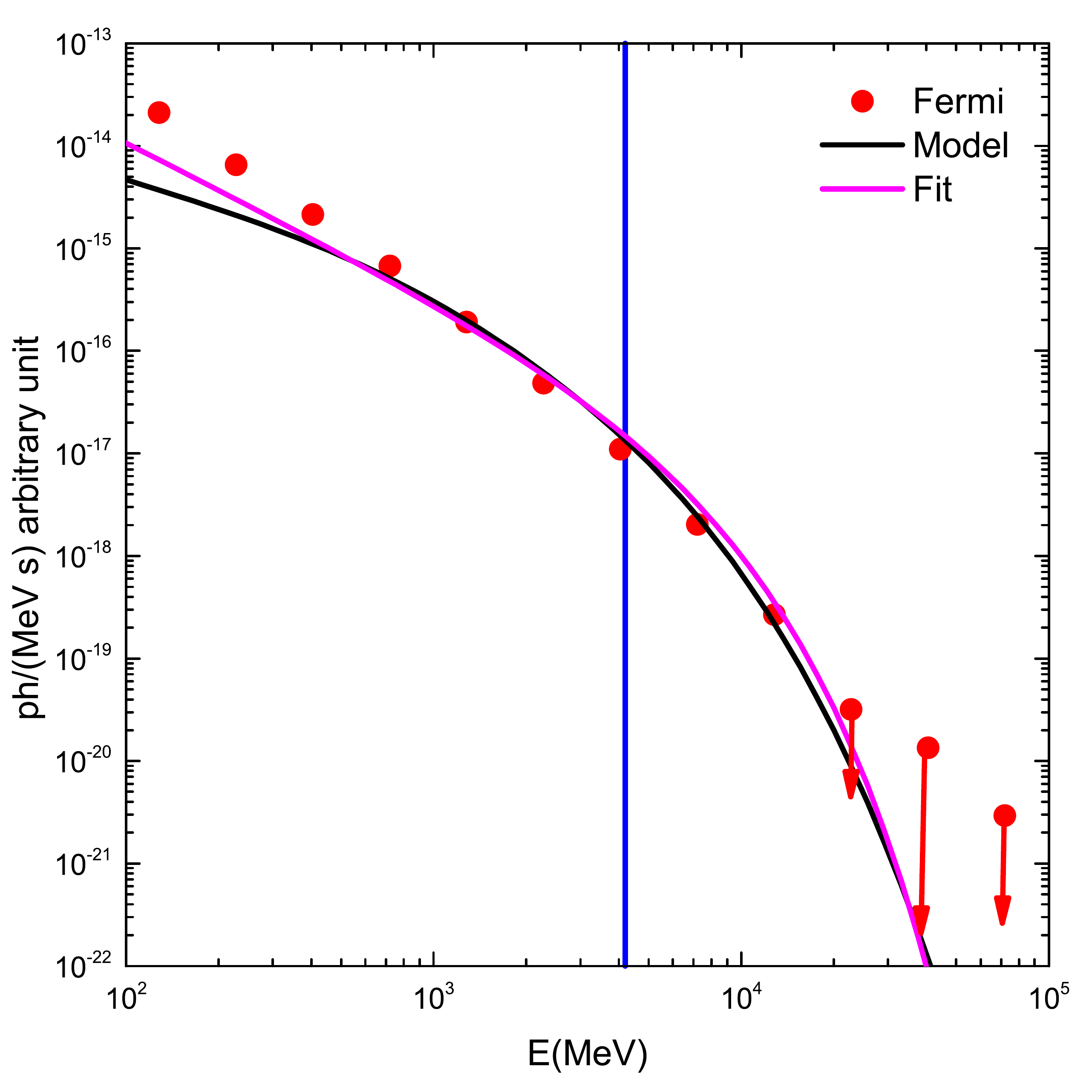}\\
\includegraphics[width=5.5 cm,height=4.5 cm]{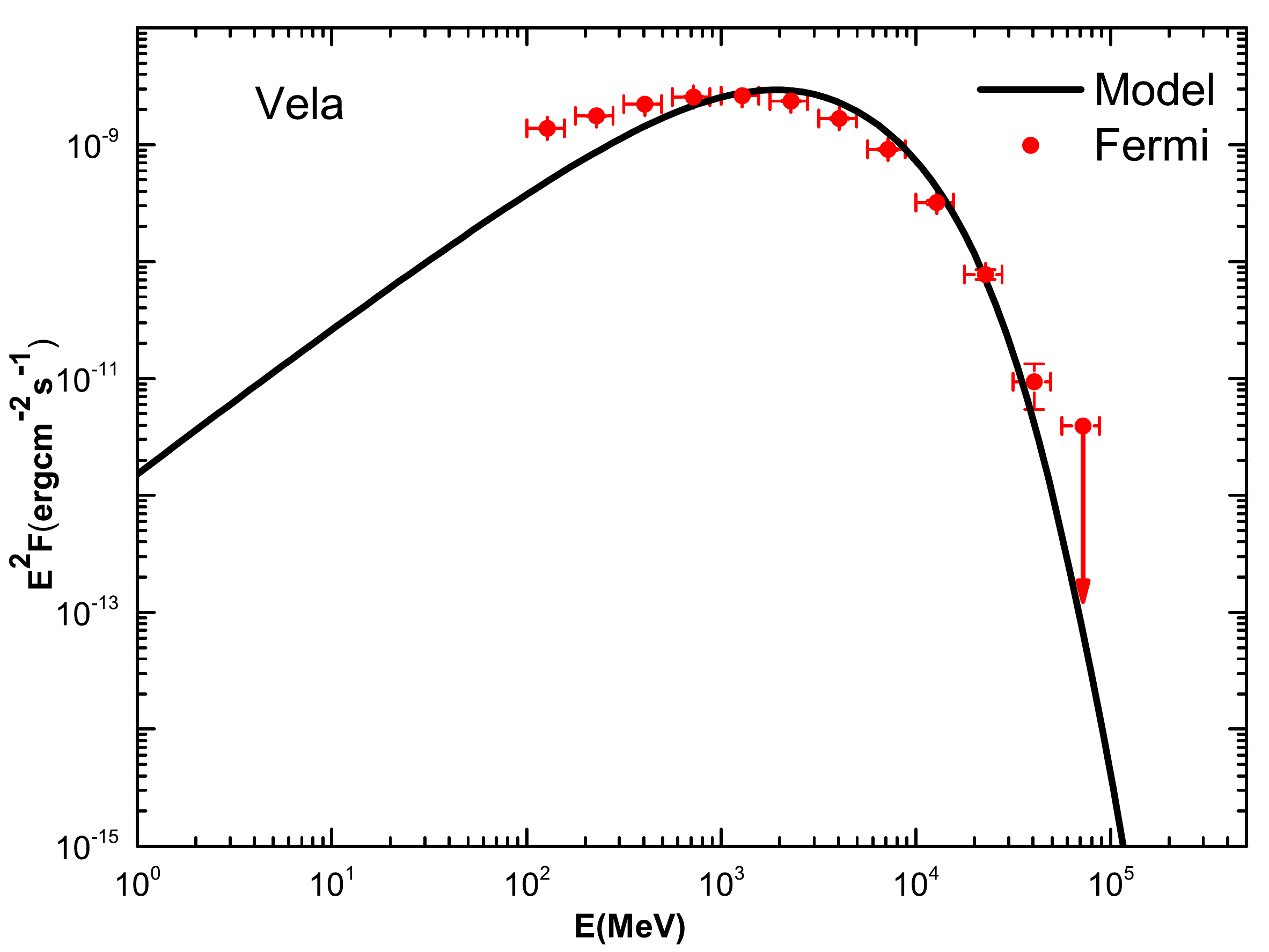}
\includegraphics[width=5.5 cm,height=4.5 cm]{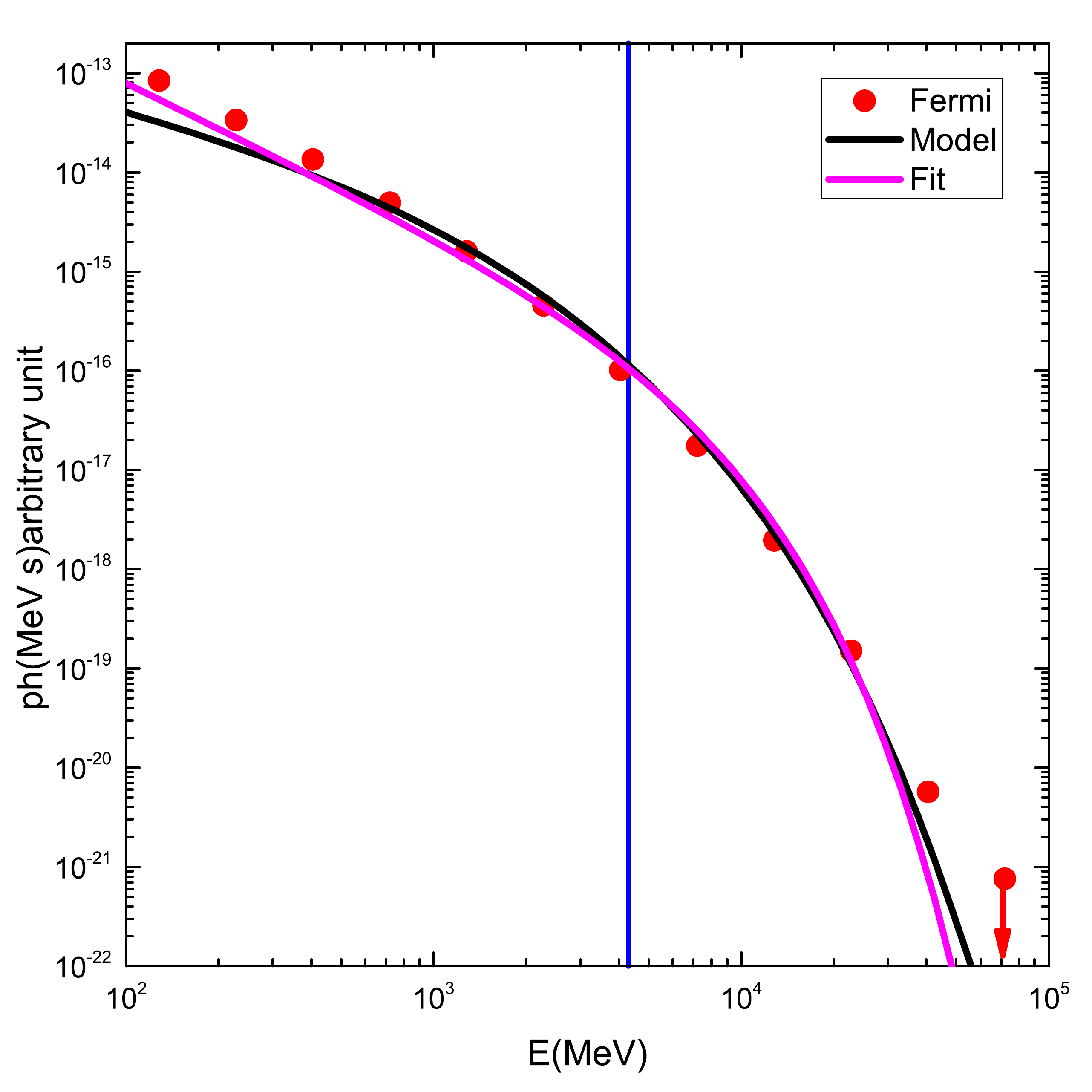}\\
\end{tabular}
\caption{The top panels are the phase-averaged(left) and differential spectra(right) for the Crab pulsar. The bottom panels are the ones for the Vela pulsar. In the left-hand column: the dark curves represent the model spectra obtained in the FIDO models. In the right-hand column: the dark curves are the differential spectra for the FIDO model, the magenta curves are the fit differential spectra of equation (\ref{dN-dE}). The red points in all the diagrams are the Fermi data. The adopted parameters are the same as in figures (\ref{fig-Crab}) and (\ref{fig-Vela}), separately. The vertical bule lines  in the differential spectra are the fit values of $E_{cut}$. For the Crab, $E_{cut}=4.2\,\rm{GeV}$. For the Vela, $E_{cut}=4.3~\rm{GeV}$.}
\label{fig-PAS}
\end{figure*}

\section{The Energy-dependent LCs and the energy spectra for the Crab and Vela pulsars}\label{sect-app}
In order to further constrain the model parameters, the emission characteristics of the Crab and Vela pulsars are studied based on our FIDO model. Their energy-dependent sky maps and light curves, phase lag $\delta$, peak width $\Delta$, phase-averaged spectra, the luminosity $L_{\gamma}$, the cutoff energy $E_{cut}$, and the spectral index $\Gamma$ are produced and compared with the ones observed by Fermi.

The model luminosity of the pulsar is determined by the following expression,
\begin{eqnarray}
L_{\gamma}=4\pi d^2 f_{\Omega} \int_{0.1}^{50}E_{\gamma}F(E_{\gamma})d{E_{\gamma}}\;,
\end{eqnarray}
where $d$ is the real distance from the pulsars from 2PC, $f_{\Omega}$ is the beam correction factor, which we choose to be one since the outer magnetosphere fan-like beam sweeping the entire sky gives $f_{\Omega}\approx 1$\citep{abd13}. $F(E_{\gamma})$ is the total energy flux in the dissipative region determined by
\begin{eqnarray}
F(E_{\gamma})=\frac{1}{\Delta\Omega\,d^2}\sum_{r} F(E_{\gamma},{\bf r}),
\end{eqnarray}
where $\Delta\Omega$ is the solid angle, which we choose to be $4\pi$ sr. The model cutoff energy $E_{cut}$ and spectral index $\Gamma$ from the phase-averaged spectrum are obtained by fitting the modeling differential spectrum with an exponential cutoff power law \citep{abd13}
\begin{eqnarray}\label{dN-dE}
\frac{dN}{dE} = K \left(\frac{E}{E_0} \right )^{-\Gamma}\,exp {\left(-\frac{E}{E_{cut}}\right)}^b\;,
\end{eqnarray}
where K is the normalization factor, $\Gamma$ the photon index, $E_{cut}$ the cutoff energy, $b$ the sharpness of the cutoff which is fixed to one in
the paper, and the energy $E_0$ at which K is defined is arbitrary.

In figures (\ref{fig-Crab}) and (\ref{fig-Vela}), we compare the predicted energy-dependent light curves with those published in Fermi 2PC for Crab and Vela pulsars,respectively. The evolutionary patterns of the energy-dependent observed light curves with the energy bands can be well explained by the FIDO models for both Crab and Vela. The dependence of the relative ratio between the two peaks on the photon energy are well reproduced by our FIDO model. We also find that the sky maps in different energy band are not changed predominantly, the similar results are also found by \citet{pet19} in the VRD magnetic fields.

Further, for the Vela pulsar, we note that the ratio of the intensity of the first peak to that of the second peak decreases as the photon energy increases. This may due to the slightly different curvature radius along the trajectories of the particles contributing to the two peaks, separately. We find that the particles contributing to the second peak will collectively possess larger $R_{\rm CR}$ than that to the first peak in the equatorial current sheet.
In the CR reaction limit, the CR cutoff energy $E_c$ is related by $E_{||}$ and the curvature radius $R_{CR}$
\begin{eqnarray}\label{eqn-Ec}
E_c\propto\,E^{3/4}_{||}R^{1/2}_{\rm CR}.
\end{eqnarray}
Even though this limit is not reached, the $E_c$ values are also expected to be larger at the second peak of the light curve than that of the first peak, which will give larger cutoff energy for the second peak than that of the first peak, consistent with the observed data \citep{de11}.
Similar results are also found by \citet{bra15} in the similar FIDO models and by \citet{bar17} in the 3D FF magnetospheres using the SG model.
Moreover, \citet{kal17} even explored the dependence of the $E_c$ on the spin down rate, which put more constraints on the FIDO models.

In figures (\ref{fig-PAS}), we plot the phase-averaged spectra and the corresponding model differential spectra are fitted to obtain the photon index and the cutoff energy for Crab (top row) and Vela (bottom row) pulsars, respectively. We see that the model CR phase-averaged spectra are well consistent with the Fermi data. Moreover, the shapes of the model spectra are very similar to the ones obtained by \citet{har15} and \citet{pet19} in FF and VRD magnetic fields, separately. The parameters adopted to produce the modeling results for the FIDO models and the ones observed by Fermi are listed in the Appendix. From which we can see that our FIDO models can better produce the results that are consistent with the ones observed. We note that the model results, i.e., the evolutionary patterns of the light curves with energies, the cutoff energy, and the spectral index for the Crab and Vela pulsars, are consistent with those obtained by \citet{bra15}.

\section{Conclusions and Discussions  }\label{sect-CD}
In this paper, we study the pulsar $\gamma$-ray light curves and  spectra based on the FIDO magnetospheres. The FIDO magnetospheres with near force-free regime inside the LC and finite but high conductivity outside the LC are constructed by a spectral algorithm. We expand the study of \citet{cao19} by computing the curvature radiation spectrum. The realistic particle trajectory is defined by using the FIDO field structures. The particle Lorentz factors along each trajectory are computed under the effects of the accelerating electric field and the curvature radiation losses. The $\gamma$-ray sky maps and light curves are then produced by collecting the curvature photons from all the emitting particles. Our results show that the distributions of $\gamma_{\rm max}$ are asymmetric around the PC edges and all the  $\gamma_{\rm max}$ values are almost concentrated on the leading side of PC edges, and that the higher Lorentz factors are coming from the particles whose trajectories are reaching near the equatorial current sheet outside the LC. The asymmetries of the high $\gamma_{\rm max}$ values around the PC also indicate that the effects of the equatorial current sheet on the particles are not symmetrical. As an application, we compare the predicted light curves and energy spectra with those of the Crab and Vela pulsars observed by Fermi. We find that the light curves and energy spectra from Crab and Vela pulsars can be well reproduced by the FIDO model. For Vela pulsar, the relative ratio of the first peak to the second peak decreases with increasing the energy. This may due to the geometrical differences of the trajectories of the particles contributing to the two peaks, respectively.

In fact, the origin of the conductivity parameter $\sigma$ is yet not completely understood, it is usually assumed that $\sigma$ is constant in some region of the magnetosphere, which maybe a strong constraint. A possible case is that $\sigma$ is a function of the distance from the neutron star \citep{Kato2017}. Therefore, it is worthy for studying the feature of the conductivity parameter in the future.
We also note that there are no back-reaction of photons onto radiative particles in the resistive model. A better approximation is radiation reaction limit (called Aristotelian Electrodynamics), where particle acceleration is fully balanced by radiation. The dissipative pulsar magnetospheres with radiation reaction limit have been presented in our another paper by \citet{cao20}. It is found that the accelerating electric field is restricted to the current sheet and the accelerating region is self-consistently controlled by the pair multiplicity. In the next step, we will use the dissipative magnetosphere with radiation reaction limit to explore the the influence of the pair multiplicity on the pulsar $\gamma$-ray emission.

It is also noted that the current numerical simulation still can not resolve  the realistic ratios of stellar to light cylinder radius. A large ratio with \mbox{$R_{*}/R_{\rm L}=0.2$} corresponding to a 1 ms pulsar is used to model the light curves and energy spectra for all pulsars \citep{kal14,bra15,cao19}. The spectral algorithm can allow us look deeply into the magnetosphere with physically realistic ratios of stellar to light cylinder radius,  and the effect of the ratios of stellar to light cylinder radius on the pulsar magnetosphere will be explored with higher resolution simulation by the spectral algorithm in our future work.

\acknowledgments
We would like to express our gratitude to the anonymous referee for the valuable comments and suggestions, and to Xuening Bai and Li Zhang for some useful discussions. This work is partially supported by National Key R \& D Program of China under grant No. 2018YFA0404204, and the National Natural Science Foundation of China U1738211. XY is also supported by Research Innovation Fund of Yunnan University 2019Z035. GC is supported by the National Natural Science Foundation of China No.12003026. The supercomputer resources supporting the work are provided by HPC Center of Yunnan University.

\appendix
Below, we list the comparison between the Fermi observed parameters and those given by our FIDO model for the Crab and Vela pulsar. The units for the period $P$, magnetic field $B$, distance $d$, luminosity $L_{\gamma}$, and the cutoff energy $E_{cut}$ are second ($\rm s$), Gauss ($\rm G$), $\rm {kpc}$, $ \rm {erg \cdot {cm}^{-2}{s}^{-1}}$, and $\rm {GeV}$, respectively.\\

\section{Crab pulsar}
(\uppercase\expandafter{\romannumeral1}). The Fermi observed parameters: $P=0.033$, $B=3.8\times10^{12}$, $d=2.0$, $E_{cut}=4.2$, $\Gamma=1.9$, $L_{\gamma}=6.2\times10^{35}$, $\delta=0.12$, and $\Delta=0.40$.

(\uppercase\expandafter{\romannumeral2}). The model-adopted parameters: $E_{cut}=4.2$, $\Gamma=1.5$, $L_{\gamma}=1.31 \times 10^{35}$, $\delta=0.12$, and $\Delta=0.45$.

Besides, the geometrical parameters used in the models  are $P=0.033$, $B=4\times10^{12}$, $d=2.0$, magnetic inclination angle $\alpha=75^\circ$, $\sigma=10$ ,and viewing angle $\zeta=50^\circ$.

\section{Vela pulsar}
(\uppercase\expandafter{\romannumeral1}). The Fermi observed parameters: $P=0.089$, $B=4\times10^{12}$, $d=0.29$, $E_{cut}=3.0$, $\Gamma=1.5$, $L_{\gamma}=8.9\times10^{34}$, $\delta=0.14$, and $\Delta=0.43$.

(\uppercase\expandafter{\romannumeral2}). The model-adopted parameters: $E_{cut}=4.3$, $\Gamma=1.5$, $L_{\gamma}=6.9 \times 10^{34}$, $\delta=0.15$, and $\Delta=0.40$.

Besides, the geometrical parameters used in the models  are $P=0.1$, $B=4\times10^{12}$, $d=0.29$, magnetic inclination angle $\alpha=60^\circ$, $\sigma=30$ and viewing angle $\zeta=42^\circ$.


\end{document}